\documentclass[runningheads]{llncs}
\usepackage{graphicx}
\usepackage[colorlinks=True]{hyperref}
\usepackage{amsmath}
\usepackage{amssymb}
\usepackage{wasysym} 
\usepackage{soul} 
\usepackage{xcolor}
\usepackage{cancel} 
\usepackage{lscape}
\usepackage[left=0.7in,right=0.7in,top=1in,bottom=0.7in]{geometry} 
\usepackage{upgreek} 
\usepackage{booktabs} 
\usepackage{tabularx} 
\usepackage{outlines} 
\usepackage{enumitem}
\setlist{topsep=1em, itemsep=0.5em} 

\usepackage{nicematrix}

\begin{document}
\title{
The hidden architecture of connections: How do multidimensional identities shape our social networks?}
\author{Samuel Martin-Gutierrez\inst{1} \and%
Mauritz N. Cartier van Dissel\inst{1}\inst{2} \and%
Fariba Karimi\inst{2}\inst{1}}
\authorrunning{S. Martin-Gutierrez}
%
\institute{Network Inequality Group, Complexity Science Hub, Josefstaedter Str. 39, 1080 Vienna, Austria
\and
Institute of Interactive Systems and Data Science, Graz University of Technology, Sandgasse 36, 8010 Graz, Austria}
\maketitle              
\begin{abstract}
Our multidimensional identities determine how we interact with each other, shaping social networks through group-based connection preferences. While interactions along single dimensions have been extensively studied, the dynamics driving multidimensional connection preferences remain largely unexplored. In this work, we develop a network model of multidimensional social interactions to tackle two crucial questions: What is the structure of our latent connection preferences, and how do we integrate information from our multidimensional identities to connect with others? To answer these questions, we systematically model different latent preference structures and preference aggregation mechanisms. Then, we compare them using Bayesian model selection by fitting empirical data from high school friendship networks. We find that a simple 
latent preference model
consistently outperforms more complex alternatives. The calibrated model provides robust measures of latent connection preferences in real-world networks, bringing insights into how one- and multidimensional groups interact. Finally, we develop natural operationalizations of dimension salience, revealing which aspects of identity are most relevant for individuals when forming connections.
\end{abstract}
%
%
%


\section{Introduction}

Social categories, such as race, gender, or socioeconomic status, are core driving forces of social tie formation. They shape our identities, determining our social behavior and our connection preferences \cite{meadMindSelfSociety2015}. 
Although our identities, and therefore our interactions, are multidimensional, much of the literature has adopted a one-dimensional perspective, studying interactions along single attributes \cite{currariniIdentifyingRolesRacebased2010a,mollicaRacialHomophilyIts2003} or considering several attributes independently \cite{newmanStructureInferenceAnnotated2016,marsdenHomogeneityConfidingRelations1988}. Early works from social psychology recognized the inherent multidimensionality of our identity, but postulated the existence of a salient dimension that would control social behavior in each context \cite{tajfelSocialIdentityIntergroup2010,ContemporarySocialPsychological2018,mollicaRacialHomophilyIts2003}. Later developments have raised the question of when and how people occupy multiple identities simultaneously \cite{smith-lovinSelfIdentityInteraction2003} 
or, equivalently, when are multiple identities simultaneously salient \cite{ContemporarySocialPsychological2018}. 

In recent years, multidimensional identities have been increasingly incorporated into the study of social networks \cite{campigottoSchoolFriendshipNetworks2022,blockMultidimensionalHomophilyFriendship2014,schaefer2010configurational,wimmerRacialHomophilyERG2010}. However, a major limitation of these studies is that they make implicit assumptions about how connection preferences are structured. For example, some assume that social groups only have two possible tendencies, in-group and out-group \cite{wimmerRacialHomophilyERG2010}, lumping together different preferences.
Another common assumption is that connections based on numerical or ordinal dimensions like age or socioeconomic status (SES) are solely driven by similarity \cite{blockMultidimensionalHomophilyFriendship2014}. As a consequence, they might assume that individuals of high SES are equally likely to connect with those of low SES as vice versa, when in reality these preferences are often asymmetric.
These assumptions, while sometimes informed by theoretical arguments, are not always justified and can obscure and occasionally distort the analysis of connection preferences. Thus, if we are to understand how our multidimensional identities determine our social connections, we first need to address the open question of how we internally represent our connection preferences depending on our group memberships. However, the problem does not end there, as we also need to identify the aggregation mechanisms we use to integrate multidimensional information and articulate our preferences into connection choices. 
While there is a severe scarcity of systematic social network studies about how we aggregate information from several dimensions when choosing with whom to connect \cite{dimaggioNetworkEffectsSocial2012,blockMultidimensionalHomophilyFriendship2014}, research on social categorization has cleared part of the path through survey-based studies that ask egos to rate alters according to a collection of attributes \cite{migdalEffectsCrossedCategorization1998,vanoudenhovenAdditiveInteractiveModels2000,uradaCrossedCategorizationTwogroup2007,grigoryanCrossedCategorizationOutside2020}.

In this work, we develop a framework for modeling multidimensional interactions and use it as a test bed to empirically validate preference structures and aggregation mechanisms. We then apply it to unravel latent connection preferences in real-world social networks. By examining the dependencies between preferences and nodes' attributes, we provide insight into the internal processes we use to represent and integrate information from multiple attributes when we interact with each other. 

The first matter we need to consider when approaching multidimensional interactions is attribute correlation. In his seminal work on intergroup relations, Blau discussed the substantial effects of attribute correlation on connectivity patterns \cite{blauInequalityHeterogeneityPrimitive1977}. For example, if socioeconomic status (SES) and race are correlated, high homophily in SES will necessarily result in high rates of homophilous association by race, even if people's preferences were neutral in that dimension \cite{moodyRaceSchoolIntegration2001}. Correlation thus heavily impacts connection patterns, complicating the measurement of connection preferences because preferences measured in a given dimension might be caused by connection tendencies in another
\cite{santiagoExtendedFormalismPreferential2008,leszczenskyEthnicSegregationFriendship2015,kronebergWhenEthnicityGender2021}. 
We control for attribute correlation by explicitly considering the population distribution in our interaction model, which in turn controls for the pool of opportunities to meet people from different groups \cite{mcphersonBirdsFeatherHomophily2001,feldFocusedOrganizationSocial1981}.

The second issue we must address is the structure of latent preferences, or the dependence of preferences on group membership. In the example above, do preferences in the race dimension depend on the SES dimension, or are they independent? For instance, do people with high SES have the same connection preferences for people from every race? Do they have the same or different connection preferences as people from lower SES? This aspect of multidimensional interactions has been only superficially explored in the literature. Most works focus on homophily, as early results showed that the strongest preferences were those associated with same-group versus different-group labels, regardless of the specific categories \cite{marsdenHomogeneityConfidingRelations1988}. In this work, we consider a varied collection of preference structures, avoiding oversimplification to identify all the relevant preference trends.

The final component of the model is a mechanism to aggregate multidimensional preferences. The necessity for a systematic characterization of aggregation mechanisms has been stressed by DiMaggio and Garip \cite{dimaggioNetworkEffectsSocial2012}, who illustrate it with an example of the possible association preferences of a female Hispanic college graduate. Would she desire to maintain friendships with people identical to herself on all three dimensions so that, in effect, female Hispanic college graduate acts as a single category? Would she connect with others based on a (possibly weighted) average of those three attributes? To answer these questions, we have devised a flexible multidimensional interaction model so that any such aggregation mechanism can be explicitly modeled, and hypotheses about how preferences are structured can be tested. 

We derive closed-form expressions for the likelihood function of the model and fit real-world data of high school friendship networks. Then, we use Bayesian model selection to compare latent preferences and aggregation mechanisms, revealing which is the most likely to describe our interaction dynamics. Finally, we illustrate the applicability of the model to study inter-group multidimensional connection preferences.

\section{Network model of multidimensional group interactions}

Let us consider a directed network where each node $i$ belongs to a group defined by a multidimensional vector of attributes $\mathbf{s}\in \mathcal{S}=S_1\times S_2\times \dots \times S_D=\{1,\dots,v_1\}\times \{1,\dots,v_2\}\times \dots \times \{1,\dots,v_D\}$, so that we have $D$ dimensions and $s^d \in \{1,\dots,v_d\}$ is the value of $i$'s attribute in dimension $d$, which has $v_d$ different possible values. Therefore, there are $\mathcal{G}=\prod_{d=1}^{D} v_d$ multidimensional groups. 
For example, in a context where sex and nationality are the relevant dimensions, with sex having two possible values (female - $\female$ and male - $\mars$) and nationality three ((A)rmenian, (B)razilian, and (C)hinese), the attribute vectors describing multidimensional groups would be $\mathbf{s} = \{(\female, A), (\female, B),(\female, C), (\male,A), (\male, B), (\male, C) \}$. Here, a multidimensional group is any fraction of the total population defined by a set of attributes.
We encode the population distribution in a tensor $F$ where the element $F_{s^1,s^2,\dots,s^D}$ is the fraction of individuals in the multidimensional group $\mathbf{s}=(s^1, s^2\dots,s^D)$ and $\sum_{s^1,s^2,\dots,s^D} F_{s^1,s^2,\dots,s^D} = 1$.

\begin{figure}
\centering
\includegraphics[width=1.0\textwidth]{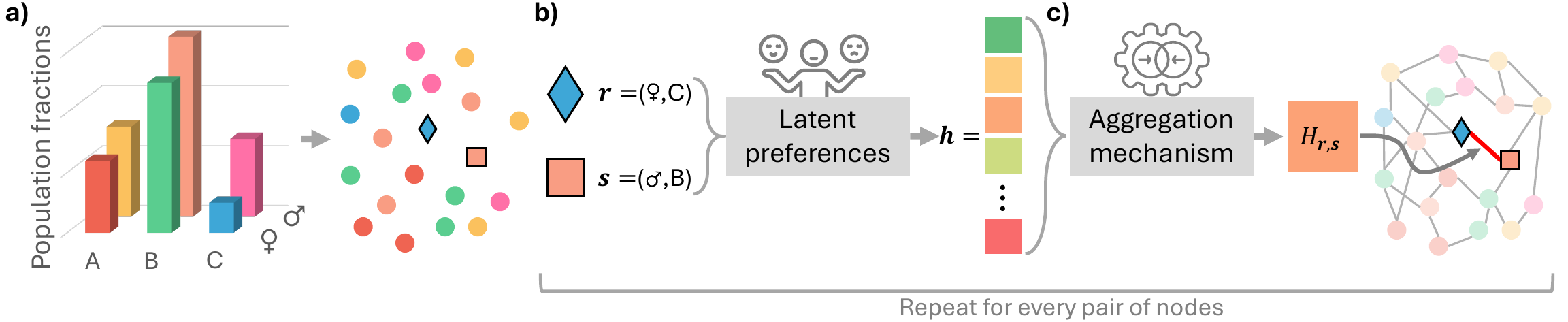}
\caption{\textbf{Diagram of the multidimensional network model}. Network formation dynamics in a two-dimensional system where the relevant dimensions are sex (female $\female$ or male $\mars$) and nationality ((A)rmenian, (B)razilian, (C)hinese). In panel \textbf{a}, we generate a set of $N$ nodes sampling from the population distribution and pick a pair of nodes, one belonging to group $\mathbf{r}$hombuses and another to group $\mathbf{s}$quares. In panel \textbf{b}, we build group $\mathbf{r}$'s latent preference vector $\mathbf{h}$ for group $\mathbf{s}$. In panel \textbf{c}, the preferences are combined through an aggregation mechanism, leading to a tie formation probability $H_{\mathbf{r},\mathbf{s}}$ that determines whether a link is established between the two nodes.}
\label{fig:model}
\end{figure}

The nodes also have some latent connection preferences that depend on their group memberships and act as \emph{partial connection probabilities}. We will discuss the form of these preferences below.
To generate a network, we consider a fixed set of $N$ nodes with attribute vectors picked from the population distribution $F$, as illustrated in panel \textbf{a} of Fig.~\ref{fig:model}.
Then, we pick a random pair of nodes $i$ and $j$; let us assume that node $i$ belongs to multidimensional group $\mathbf{r}$  (rhombuses) and node $j$ belongs to multidimensional group $\mathbf{s}$ (squares). The nodes' multidimensional groups determine $i$'s \emph{latent preferences} to connect to $j$, which can be encoded in a vector $\mathbf{h}$ (see Fig.~\ref{fig:model}\textbf{b}). 
To decide whether or not to establish a connection, $i$ uses an \emph{aggregation mechanism} to combine its latent preferences into a single quantity $H_{\mathbf{r},\mathbf{s}}$ that corresponds to the final tie formation probability, as in Fig.~\ref{fig:model}\textbf{c}.

We call the multidimensional preference matrix $H_{\mathbf{r},\mathbf{s}}$  because the diagonal values measure \emph{homophily}, or in-group preference, and the off-diagonal values measure \emph{heterophily}, or out-group preferences. 
{While $\mathbf{r}$ and $\mathbf{s}$ are categorical vectors, to build the $H_{\mathbf{r},\mathbf{s}}$ matrix we map them to numerical indices using a one-to-one mapping defined in Methods. The 6 vectors enumerated in the example above would each be mapped to an index between 1 and 6. Once the correspondence between one-dimensional categorical labels (male, Brazilian) and indices $(2,2)$ is fixed, the mapping between vectors and their indices is unique $(2, 2)\rightarrow 5$.}

We make a tie-formation attempt for each of the $N(N-1)$ possible node pairs, so our approach can be considered a multidimensional version of the Stochastic Block Model (SBM) \cite{newmanNetworksIntroduction2010}. However, the core elements of our model are
the latent connection preferences and aggregation mechanisms that map attribute vectors to connection probabilities. These connection probabilities can be used to create links with any other stochastic tie formation mechanisms. For example, we could add one node at a time, as in network growth models \cite{karimi_homophily_2018}, or we could rewire an existing network \cite{asikainenCumulativeEffectsTriadic2020}.

The multidimensional preference matrix  $H_{\mathbf{r},\mathbf{s}}$ is determined by two factors: the dependence of \emph{latent preferences} on group memberships and the \emph{aggregation mechanisms} we (unconsciously) use to combine several preferences and decide whether or not to establish a tie. Determining the latent preferences and aggregation mechanisms that shape $H_{\mathbf{r},\mathbf{s}}$ is one of our central objectives.

\subsection{Modeling latent connection preferences}

\begin{figure}
\centering
\includegraphics[width=1.0\textwidth]{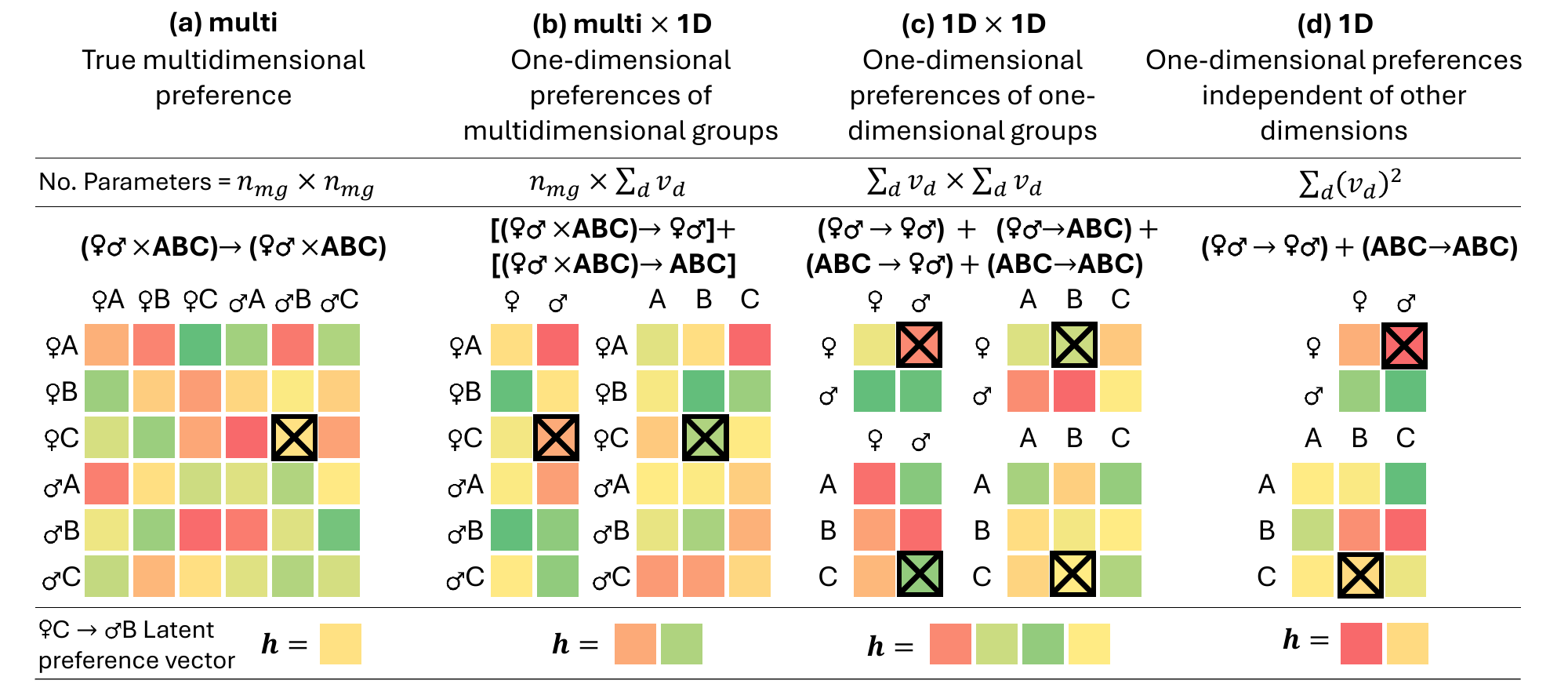}
\caption{\textbf{Latent preference structures}. Schematic preference matrices illustrating the possible ways in which preferences may depend on multidimensional or one-dimensional group membership in a two-dimensional system where the relevant dimensions are sex (female $\female$ or male $\mars$) and nationality ((A)rmenian, (B)razilian, (C)hinese). Each element of the matrices represent the preference of the group in the row to connect to the group in the column. }
\label{fig:preference_dependencies}
\end{figure}

Regarding how preferences may depend on group membership, we can consider two extreme cases. On one extreme, preferences may be fully multidimensional, depending on the source and target nodes' multidimensional group membership (see Fig.~\ref{fig:preference_dependencies}\textbf{a}). On the other extreme, preferences may be fundamentally one-dimensional and independent across dimensions, so that preferences in dimension $d$ are solely determined by the group memberships of the source and target nodes in that dimension, and are independent of the nodes' categories in any other dimension (see Fig.~\ref{fig:preference_dependencies}\textbf{d}). 
In between these extremes, preferences might depend on the source node's multidimensional group and the target node's one-dimensional categories (Fig.~\ref{fig:preference_dependencies}\textbf{b}), or on one-dimensional group memberships, including cross-dimensional preference (Fig.~\ref{fig:preference_dependencies}\textbf{c}). In Fig.~\ref{fig:preference_dependencies}, we represent the possible latent preference structures discussed here as preference matrices of the two-dimensional system described above, where sex and nationality are the relevant dimensions.

For illustration, we also mark the preferences for the interaction between a Chinese woman ($\female,C$) and a Brazilian man ($\male,B$).
If preferences depend on the multidimensional group membership of the source and the target nodes, we need to know every element of $H_{\mathbf{r},\mathbf{s}}$ to describe the preference structure completely.
In our example, Chinese women would have 6 distinct preferences for each of the multidimensional groups, and the model would have $\mathcal{G} \times \mathcal{G}=6\times6=36$ parameters (Fig.~\ref{fig:preference_dependencies}\textbf{a}).
If preferences depend on the source node's multidimensional group but only differ between the one-dimensional groups of target nodes, Chinese women have a certain preference for each gender and for each race, which they then combine into a multidimensional preference using some aggregation mechanism.
In this case, there would be $\mathcal{G}\sum_d v_d = 6 \times 5 = 30$ parameters that we arrange in $D$ preference matrices of size $\mathcal{G}\times v_d$  (two matrices of size $6\times2$ and $6\times 3$ in our example of Fig.~\ref{fig:preference_dependencies}\textbf{b}).
A further simplification would consider preferences dependent on the one-dimensional groups of source and target nodes. Thus, Chinese people would have certain preferences for each gender and each nationality, women would independently have certain preferences for the different genders and nationalities, and Chinese women would aggregate the corresponding preferences due to being Chinese and a woman into a compound preference for each multidimensional group. This model has $\left(\sum_d v_d \right)^2 = 25$ parameters in our example, which we arrange in $D^2$ matrices of size $v_{d_1} \times v_{d_2} : d_1,d_2 \in \{1,\dots, D\}$. Therefore, we would have four matrices of sizes $2\times2$, $2\times3$, $3\times2$, and $3\times3$ (Fig.~\ref{fig:preference_dependencies}\textbf{c}).
Finally, preferences may depend only on the one-dimensional groups of the focal dimension, such that the preferences of Chinese women for men or women only depend on them being women, and their preferences for other nationalities depend on them being Chinese. In this case, there are $\sum_d v_d^2 = 13$ free parameters that we arrange in $d$ latent preference matrices of size $v_d \times v_d$ (Fig.~\ref{fig:preference_dependencies}\textbf{d}).

The task of determining the preference structure that shapes our interactions is closely linked to theoretical discussions and empirical evidence from social psychology, which can help us build theoretical foundations about how these preferences may manifest themselves. Several theories of multidimensional identity explore this question by focusing on the notion of \emph{identity salience}. An identity dimension is salient when it is relevant to the formation of connections. Although a hierarchical structure of dimensions in terms of salience has been hypothesized, there is no quantitative definition for this concept, and the problem of determining when multiple identities are simultaneously salient remains an open question \cite{ContemporarySocialPsychological2018}. With our framework, we can measure salience as a departure from neutral preference. Salient dimensions will show strong preferences (very high and very low values of $h$), while irrelevant dimensions will have weak preferences (moderate and relatively similar values of $h$). According to social psychological theories, we can expect a hierarchy of salience across dimensions that changes depending on the context. In Section \ref{sec:empirical_onedim_prefs} we explore two operationalizations of dimension salience in real-world networks.

Except for full multidimensional preferences (Fig.~\ref{fig:preference_dependencies}\textbf{a}), in all the situations described above a source node needs to aggregate several preferences to decide whether it will create a link with the candidate target node. For example, in  Fig.~\ref{fig:preference_dependencies}\textbf{d}, an individual has to aggregate two preference values: one for gender and one for nationality. 

\subsection{Modeling preference aggregation mechanisms}
\label{sec:aggregation_functions}

Finding the information aggregation mechanism people use in interactions is a long-standing open question that has been formulated in different ways. For instance, DiMaggio and Garip \cite{dimaggioNetworkEffectsSocial2012} formulate the question by asking how a person (for example, a female Hispanic college graduate) who prefers to associate with people like herself in several dimensions (e.g. sex, ethnicity, and academic attainment) would articulate that preference. Would she...
\begin{enumerate}
    \item ... attempt to maintain friendships with people identical to herself on all three dimensions so that her multidimensional group membership acts as a single category? This would correspond to the full multidimensional case discussed above and would not require an aggregation mechanism. 
    \item ... use a weighted average of those preferences? 
    \item ... make friends in each of the three categories independently? 
    \item ... follow an exclusion principle, looking for the maximum similarity but avoiding alters in a particular category (e.g. high-school dropouts)?
\end{enumerate}

In our model, we consider latent preferences as \emph{partial probabilities} of establishing a tie. This probabilistic perspective enables the operationalization of aggregation functions by transparently describing microscopic decision mechanisms. Below, we discuss three mechanisms and their associated aggregation functions using the 1D preference structure of Fig.~\ref{fig:preference_dependencies}\textbf{d}, where preferences are one-dimensional and independent of other dimensions. Therefore, we arrange preferences in $D$ latent preference matrices of size $v_d \times v_d$ that we call $h^d$. The \emph{partial probability} for establishing a tie from category $r^d$ to category $s^d$ in dimension $d$   would be the matrix element $h^d_{r^d, s^d}$. Preferences organized according to the structures shown in panels \textbf{b} and \textbf{c} of Fig.~\ref{fig:preference_dependencies} can be similarly aggregated using these mechanisms. The mechanisms are the following:

\paragraph{\textbf{Average preference (\emph{MEAN}).}}

Multidimensional preference may be a weighted average of one-dimensional preferences (as in point 2 of the discussion above):

\begin{equation}
    H_{\mathbf{r},\mathbf{s}} = \sum_{d=1}^D w_d h^d_{r^d,s^d}
    \label{eq:weighted_average}
\end{equation}

Where the weight vector ($\sum_d {w_d} = 1$) can be used to encode the allocation of (limited) cognitive resources into a subset of prioritized or salient attributes \cite{tamaritCognitiveResourceAllocation2018}. Given its functional form, we call this aggregation function \textbf{MEAN}.

From a network perspective, this kind of aggregation mechanism models a situation where two people meet and only evaluate one random dimension $d$, making a link depending on the corresponding one-dimensional affinity value $h^d_{r^d,s^d}$. If all the elements of the weight vector $\mathbf{w}$ are the same, the dimension of interaction is chosen uniformly at random. To derive Eq. \eqref{eq:weighted_average} from this dynamics, we interpret $w_d$ as the probability that a pair of nodes interact through dimension $d$ when they meet.
Then, the joint probability of first choosing dimension $d$ to interact and then successfully establishing a link is
$w_d h^d_{r^d,s^d}$. Finally, to get the total probability of making a link from a node of group $\mathbf{r}$ to a node of group $\mathbf{s}$, we sum over all dimensions, obtaining Eq. \eqref{eq:weighted_average}.


\paragraph{\textbf{Preference in all dimensions (\emph{AND}).}}

Inflexible or \emph{picky} individuals may only establish ties with people with whom they have high affinity in all dimensions. Such an individual would evaluate each dimension independently, establishing a link only if all the one-dimensional assessments are positive.  
This is equivalent to attempting a connection in every dimension, each with probability $h^d_{r^d,s^d}$, and considering the overall interaction successful only if all attempts are successful. Since the attempt in each dimension is independent of the others, the resulting multidimensional preference is:

\begin{equation}
    H_{\mathbf{r},\mathbf{s}}  = \prod_{d=1}^D h^d_{r^d,s^d}
    \label{eq:all_dim}
\end{equation}

This aggregation mechanism can be interpreted as a simple composition of independent preferences: we have $D$ independent preferences encoded as partial probabilities of tie formation, and the probability of finally making the tie is the probability of success of these $D$ independent coin tosses (each with success probability $h^d_{r^d,s^d}$). This operation is reminiscent of the logical operation \emph{and}, where the output is 1 only if all the inputs are 1 (although $h$ can take values between 0 and 1), so we have called it \textbf{AND}.

The \textbf{AND} mechanism is one of the possible operationalizations of the independent aggregation of preference described in point 3 of the example above, but it can also model an exclusion principle (point 4) by simply setting the $h^d_{r^d, s^d}$ preference for the excluded dimension (high-school dropouts in the example) to 0.

\paragraph{\textbf{Preference in any dimension (\emph{OR}).}}

In the other extreme, people may be satisfied with finding affinity in any dimension, in which case the final probability of connection is:

\begin{equation}
    H_{\mathbf{r},\mathbf{s}} = 1 - \prod_{d=1}^D (1-h^d_{r^d, s^d})
    \label{eq:at_least_one_dim}
\end{equation}

The mechanism in this case would be one where two people meet, assess all dimensions, and a single positive evaluation in one dimension is enough to connect. Following the analogy with logical operations, we have called this aggregation function \textbf{OR}. This is another way in which the mechanism described in point 3 could work. 

\begin{figure}
\centering
\includegraphics[width=1.0\textwidth]{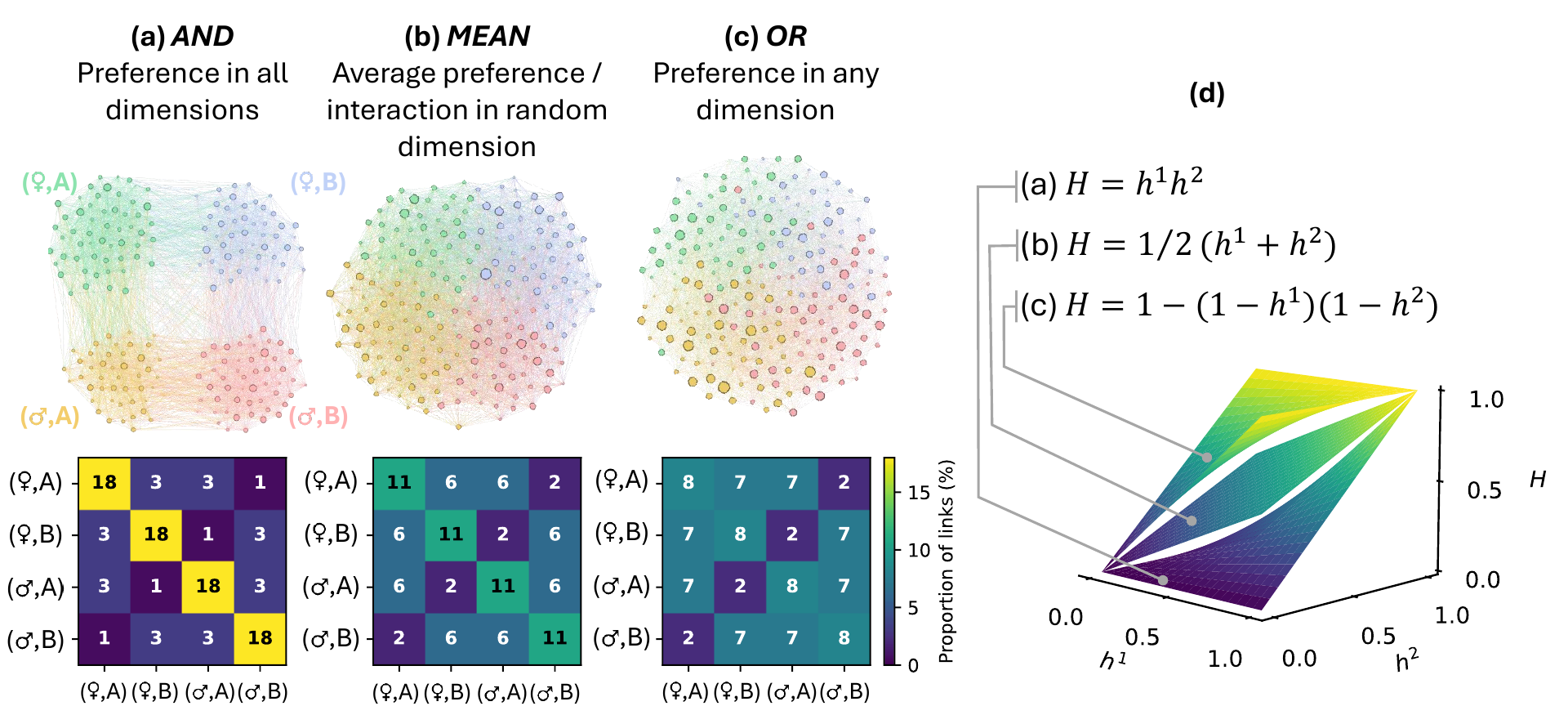}
\caption{\textbf{Aggregation mechanisms}. Panels \textbf{a-c} show synthetic networks of a two-dimensional system with two categories per dimension ($\female, \male$ and A, B) generated with different aggregation functions (top row) and the proportion of links between each multidimensional group (bottom row). Panel \textbf{d} shows the resulting multidimensional homophily $H$ associated with each of the considered aggregation mechanisms for any pair of one-dimensional preference values (or partial tie formation probabilities) $(h^1,h^2)$. Color scales with the value of $H$.}
\label{fig:networks_interaction_type}
\end{figure}

Notice that verbal descriptions of aggregation mechanisms are not specific enough to define aggregation functions unambiguously, so operationalizing them with aggregation functions is crucial for an accurate characterization of the interaction dynamics. In Fig.~\ref{fig:networks_interaction_type}, we show how the linking patterns of networks governed by each of these mechanisms differ for a hypothetical two-dimensional system with two categories per dimension where each of the four multidimensional groups contains 25\% of the nodes; that is, $F_{1,1}=F_{1,2}=F_{2,1}=F_{2,2}=0.25$. For this example, we continue to use the 1D preference structure of Fig.~\ref{fig:preference_dependencies}\textbf{d}, and we have set the latent one-dimensional preferences so that the two dimensions are homophilic with the same preference values. Specifically, the one-dimensional preference matrices used for all examples are:

\begin{equation}
    \begin{NiceMatrixBlock}[auto-columns-width]
    {h^1_{r^1,s^1}} = {h^2_{r^2,s^2}}= 
        \begin{bNiceMatrix}
            0.85 & 0.15 \\
           0.15 & 0.85
        \end{bNiceMatrix}
    \end{NiceMatrixBlock}
    \label{eq:1dpref_example}
\end{equation}

The aggregation that seeks high \emph{preference in all dimensions} (\textbf{AND}) results in a highly modular and segregated network (Fig.~\ref{fig:networks_interaction_type}\textbf{a}). On the other extreme, if people settle for high \emph{preference in any dimension} (\textbf{OR}), the groups mix well with each other (Fig.~\ref{fig:networks_interaction_type}\textbf{c}) because sharing one attribute is enough to have high connection probability. However, group pairs that share no attribute are still mostly disconnected. Performing an average or interacting in random dimensions (\textbf{MEAN}) leads to an intermediate level of mixing (Fig.~\ref{fig:networks_interaction_type}\textbf{b}). Panel \textbf{d} shows the values that multidimensional preference $H$ can take according to the aggregation functions associated with each of these mechanisms for an arbitrary pair of one-dimensional preferences (${h^1}, {h^2}$). This illustrates the mathematical boundaries and possibilities that each aggregation function can create, allowing us to exclude possibilities and examine plausible scenarios.

A line of research in social psychology that addresses the problem of aggregation of multidimensional identities is the study of \emph{crossed categorization}. 
In these studies, researchers run surveys where individuals are asked to express their preferences for alters with different sets of attributes by rating them. They then compare different aggregation mechanisms using statistical tests. Although there is not yet full consensus, the aggregation function with the most support is a simple additive model where preference scores from multiple dimensions are simply added up \cite{migdalEffectsCrossedCategorization1998,vanoudenhovenAdditiveInteractiveModels2000,grigoryanCrossedCategorizationOutside2020}. As we will show below, this additive model is conceptually equivalent to the aggregation function \textbf{AND}. 

Coincidentally, standard social network models such as Stochastic Actor-Oriented Models (SAOM) \cite{blockMultidimensionalHomophilyFriendship2014} and Exponential Random Graph Models (ERGM) \cite{wimmerRacialHomophilyERG2010}, at their core, implicitly assume a probability multiplication mechanism to aggregate multidimensional preferences. However, while the analyses carried out using these models offer meaningful insights, their choice of aggregation function (and preference structure) is rarely discussed and validated \cite{blockMultidimensionalHomophilyFriendship2014}. In this work, rather than taking any form of aggregation for granted, we test them against reasonable alternatives to answer this long-standing question systematically. We argue that, while making such modeling assumptions is reasonable, we should not use them blindly and they should not remain untested. They are not just specifications of a methodology, but theoretical statements about human behavior \cite{leifeldStochasticActororientedModel2022}.


In the following section, we validate our approach using Bayesian model selection to compare the performance of latent preference structures and aggregation functions on empirical data, finding ample support for the multiplicative aggregation mechanism \textbf{AND}, thus consolidating this set of independent findings from crossed categorization and other social network models. 

\section{Application to real-world social networks}

In this section, we apply our model to real-world social networks. Our goal here is twofold. First, we aim to identify the preference structure and aggregation function people use when interacting with each other by fitting datasets from different contexts and assessing which structure and function outperform the alternatives. Second, we will show the framework's usefulness for characterizing underlying connection preferences in multidimensional systems. 
However, we first need a method to infer the model parameters from data.

\subsection{Inference of model parameters}
\label{sec:model_parameters}

To infer the model parameters, we perform a maximum-likelihood fit using the network's mixing matrix $\mathbf{e}$,  whose element $e_{\mathbf{r} \mathbf{s}}$ is the number of links from multidimensional group $\mathbf{r}$ to $\mathbf{s}$. Since $\mathbf{r}$ connects to $\mathbf{s}$ with probability $H_{\mathbf{r}, \mathbf{s}}$ and there are $N_\mathbf{r}N_\mathbf{s}$ connection attempts ($N_\mathbf{r}=F_\mathbf{r}N$  is the number of nodes in group $\mathbf{r}$), we can approximate the distribution of $e_{\mathbf{r}, \mathbf{s}}$ as a binomial $B(n,p)$ of the form $e_{\mathbf{r}, \mathbf{s}} \sim B(N_\mathbf{r}(N_\mathbf{s} - \updelta_{\mathbf{r}, \mathbf{s}}),H_{\mathbf{r}, \mathbf{s}})$. Here, we have used Kronecker's delta $\updelta_{\mathbf{r}, \mathbf{s}}$ to discount the possibility of a node connecting to itself. The likelihood of the inter-group connection probability being $H_{\mathbf{r}, \mathbf{s}}$ having observed $\widetilde{e_{\mathbf{r}, \mathbf{s}} }$ links from $\mathbf{r}$ to $\mathbf{s}$ is \cite{karrerStochasticBlockmodelsCommunity2011a}:

\begin{equation}
    \mathcal{L}(H_{\mathbf{r}, \mathbf{s}} | \widetilde{ e_{\mathbf{r},\mathbf{s}} }) = {N_\mathbf{r}(N_\mathbf{s} - \updelta_{\mathbf{r}, \mathbf{s}})\choose \widetilde{ e_{\mathbf{r},\mathbf{s}} }}%
    {H_{\mathbf{r}, \mathbf{s}}}^{\widetilde{ e_{\mathbf{r},\mathbf{s}} }}%
    (1-H_{\mathbf{r}, \mathbf{s}})^{N_\mathbf{r}(N_\mathbf{s} - \updelta_{\mathbf{r}, \mathbf{s}}) - \widetilde{ e_{\mathbf{r},\mathbf{s}} }}
\end{equation}

Since the elements $e_{\mathbf{r}, \mathbf{s}}$ of the mixing matrix are independent of each other, the total likelihood of the multidimensional preference matrix $\mathbf{H}$ having observed an empirical matrix $\widetilde{\mathbf{e}}$  is the product of the binomial likelihood functions for each element:

\begin{equation}
    \mathcal{L}(\mathbf{H} | \widetilde{\mathbf{e}}) = \prod_{\mathbf{r}, \mathbf{s}} \mathcal{L}(H_{\mathbf{r}, \mathbf{s}} | \widetilde{ e_{\mathbf{r},\mathbf{s}} }) = \prod_{\mathbf{r}, \mathbf{s}}  {N_\mathbf{r}(N_\mathbf{s} - \updelta_{\mathbf{r}, \mathbf{s}})\choose \widetilde{ e_{\mathbf{r},\mathbf{s}} }}%
    {H_{\mathbf{r}, \mathbf{s}}}^{\widetilde{ e_{\mathbf{r},\mathbf{s}} }}%
    (1-H_{\mathbf{r}, \mathbf{s}})^{N_\mathbf{r}(N_\mathbf{s} - \updelta_{\mathbf{r}, \mathbf{s}}) - \widetilde{ e_{\mathbf{r},\mathbf{s}} }}
    \label{eq:total_likelihood}
\end{equation}
To recover preferences from a mixing matrix $\widetilde{\mathbf{e}}$, 
 we maximize the logarithm of this likelihood function. For multidimensional systems, we infer the underlying lower dimensional preference matrices $h$ by first replacing $H_{\mathbf{r}, \mathbf{s}}$ in Eq. \eqref{eq:total_likelihood} with its corresponding expression according to the chosen aggregation function and then finding the maximum of the log-likelihood numerically. Given the functional form of the aggregation functions, the model would be unidentifiable unless we impose some additional constraints on the $h$ matrices. In Methods Section \ref{sec:num_parameters_identifiability}, we provide a detailed discussion of this issue and how to tackle it.

If the system is one-dimensional, $\mathbf{r}$ and $\mathbf{s}$ are not vectors, but scalar indices $r$ and $s$. In that case, there is no need to consider latent preferences or aggregation functions, as $H_{r,s}$ corresponds directly to the preference $h_{r,s}$ of one-dimensional group $r$ to form ties to group $s$. The maximum likelihood estimator (MLE) of a one-dimensional system can be obtained exactly:

\begin{equation}
    \widehat{h}_{r,s} = \frac{\widetilde{e_{r,s}}}{N_r(N_s  - \updelta_{r,s})}
    \label{eq:1d_mle}
\end{equation}

Where we use lowercase $\widehat{h}_{r,s}$ to denote the MLE of the $r \rightarrow s$ preference. Notice that, if we consider the multidimensional groups as our units of analysis, we can treat the multidimensional system as a one-dimensional system with $v=\mathcal{G}$ different values. Following that argument, it is easy to see that the MLE of the multidimensional preference $\widehat{H}_{r\mathbf{,}\mathbf{s}}$ is:

\begin{equation}
    \widehat{H}_{\mathbf{r},\mathbf{s}} = \frac{\widetilde{E_{\mathbf{r},\mathbf{s}}}}{N_\mathbf{r}(N_\mathbf{s}  - \updelta_{\mathbf{r},\mathbf{s}})}
    \label{eq:multidim_mle}
\end{equation}

Where $\widetilde{E_{\mathbf{r},\mathbf{s}}}$ is the number of links from multidimensional group $\mathbf{r}$ to multidimensional group $\mathbf{s}$. 

The simplicity of the one-dimensional MLE of Eq. \eqref{eq:1d_mle} may tempt us to use it for estimating the underlying one-dimensional preferences in a multidimensional system by computing them one dimension at a time. However, this approach can lead to highly misleading results, especially if attributes are strongly correlated. To illustrate this problem, we have simulated a two-dimensional system with $N=200$ nodes and two values per dimension ($D=2,v_1=v_2=2$). We consider balanced one-dimensional population distributions $F$, such that 

$$F_{1,1}+F_{1,2} = F_{2,1}+F_{2,2} = F_{1,1}+F_{2,1} = F_{1,2}+F_{2,2} = 0.5$$

This choice constraints the population distribution up to one free parameter, which controls correlation. To fully determine $F$, we tune the correlation parameter $\kappa$ to interpolate between the population distribution with maximum values in the diagonal and the one with maximum values in the anti-diagonal:

\begin{equation}
    \begin{NiceMatrixBlock}[auto-columns-width]
    F =  \kappa
        \begin{bNiceMatrix}
        0.5 & \quad 0 \\
        0 & \quad 0.5
        \end{bNiceMatrix}
    + (1-\kappa)
        \begin{bNiceMatrix}
        0 &\quad 0.5 \\
        0.5 &\quad 0
        \end{bNiceMatrix}
        \end{NiceMatrixBlock}
\end{equation}

For this example, let us assume that preferences are structured as in Fig.~\ref{fig:preference_dependencies}\textbf{d}, with dimension 1 having homophilic preferences and dimension 2 being fully neutral; for instance:

\begin{equation}
    \begin{NiceMatrixBlock}[auto-columns-width]
    {h^1_{r^1,s^1}} = \begin{bNiceMatrix}
            0.4 & 0.1 \\
           0.2 & 0.3
        \end{bNiceMatrix} \qquad
    {h^2_{r^2,s^2}}= 
        \begin{bNiceMatrix}
            0.25 & 0.25 \\
           0.25 & 0.25
        \end{bNiceMatrix}
    \end{NiceMatrixBlock}
    \label{eq:1dpref_wrong}
\end{equation}

\begin{figure}[htb]
\centering
\includegraphics[width=1.0\textwidth]{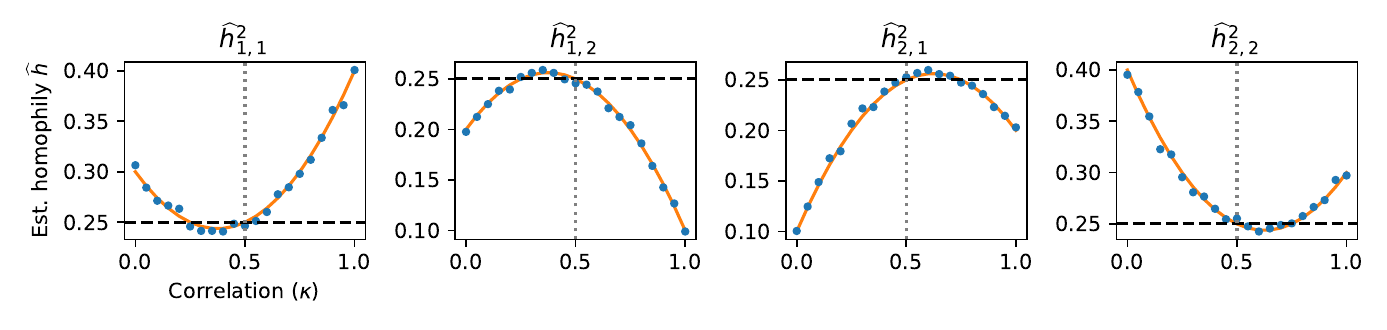}
\caption{\textbf{Misleading estimation of one-dimensional preferences}.  Estimated one-dimensional preferences in the second dimension of a two-dimensional system with actual latent preferences given by Eq. \eqref{eq:1dpref_wrong} as a function of correlation $\kappa$. Each panel shows the estimated values of an element of the preference matrix $h^2_{r^2,s^2}$. The horizontal dahsed line indicates the true preference value and the vertical dotted line, the point of no correlation or random mixing.}
\label{fig:wrong_1d_preference}
\end{figure}

We aggregate these preferences using the aggregation function \textbf{AND}. We have tuned the correlation $\kappa$ and performed 10 simulations for each value. Then, we have estimated the one-dimensional preferences of dimension $2$ with Eq. \eqref{eq:1d_mle} and we have represented the resulting estimation $\widehat{h}^2_{r^2,s^2}$ as a function of $\kappa$ in Fig.~\ref{fig:wrong_1d_preference}. The circles show the average of the 10 simulations and the lines the theoretical results, which we calculate by plugging the expected values of $e_{r^d,s^d}$ given by the model into Eq. \eqref{eq:1d_mle} (see Supplementary Section \ref{sec:biased_1d_preferences} for the details). As observed in the figure, the only point where the estimation correctly captures $\widehat{h}^2_{r^2,s^2} = 0.25$ is the one corresponding to no correlation or random mixing ($\kappa = 0.5$ in this case). 

This result demonstrates that naive estimations of one-dimensional preferences in multidimensional systems can lead to erroneous conclusions if there are strong correlations.
 Although the confounding effect of correlation on one-dimensional connection rates has long been recognized \cite{blockMultidimensionalHomophilyFriendship2014,wimmerRacialHomophilyERG2010,santiagoExtendedFormalismPreferential2008,blauInequalityHeterogeneityPrimitive1977}, the use of one-dimensional metrics that disregard the influence of correlated dimensions in multidimensional systems remains a widespread practice \cite{chettySocialCapitalMeasurement2022}. To estimate one-dimensional preferences reliably in multidimensional systems, it is imperative to explicitly model latent preferences and their aggregation. Below, we show how to use our Bayesian approach to infer latent preferences in practice regardless of how attributes are correlated in the data. The method consists in choosing a latent preference structure and aggregation function, which fully determine the functional form of $H_{\mathbf{r}, \mathbf{s}}$, plug the corresponding expression in Eq. \eqref{eq:total_likelihood}, and obtain the maximum-likelihood estimators for the underlying preference matrices $h$.

\subsection{Data description}


To validate our model with real-world networks, we use the National Longitudinal Study of Adolescent Health ({AddHealth}) dataset \cite{moody_peer_2001}, which contains information about friendship relationships among high school students in the United States and about the students' sociodemographic attributes. We use a processed version of the dataset that includes directed friendship networks and the grade (between 7th and 12th), race (black, hispanic, asian, white, and mixed/other), and gender (boy or girl) of the students \cite{peixotoAdd_healthAdolescentHealth}.

The dataset includes friendship networks from 84 \emph{communities}. Some communities include one school and others two (a middle school and a high school). We selected schools with more than 100 students remaining in the network after processing the data and with more than one gender and race category.  As a result, 41880 students from 70 communities were included in our study. We have analyzed each of the 70 communities independently. We provide all the details about the data processing and filtering in Supplementary Section \ref{sec:appendix_data}.





\subsection{Comparison of preference structures and aggregation functions}
\label{sec:test_pref_struct_aggr_fun}

To assess which of the latent preference structures of Fig.~\ref{fig:preference_dependencies} and aggregation functions of Fig.~\ref{fig:networks_interaction_type} best describe the mental representations and interaction mechanisms people use when establishing social ties, we have fitted the model with different combinations of latent preferences and aggregation mechanisms and compared these combinations using model selection metrics: the maximum log-likelihood, the Akaike Information Criterion (AIC) and the Bayesian Information Criterion (BIC). 
AIC and BIC are computed as follows:

\begin{align}
    \text{AIC} = & 2 k - 2 \log (\hat{L})\\
    \text{BIC} = & k \log(n) - 2 \log (\hat{L})
\end{align}

Where $k$ is the number of free parameters of the model (see Suppelementary Table \ref{tab:free_params}) and $n$ is the number of data points. Since we fit the model individually in each school network, $n$ corresponds to the number of elements in the mixing matrix $\widetilde{e_{\mathbf{r},\mathbf{s}}}$; that is, $\mathcal{G}\times \mathcal{G}$. Higher values of likelihood $\hat{L}$ indicate better models. However, since we work with the log-likelihood and $\hat{L} < 1$, a good model would have \emph{low} absolute values of $|\log(\hat{L})|$. For AIC and BIC, the lower the value, the better the model.

The maximum likelihood is a measure of \emph{absolute} model performance. A complex model with more parameters will almost always outperform a simpler model with fewer parameters in terms of likelihood, so optimizing likelihood alone usually leads to overfitting. To avoid overfitting, AIC and BIC penalize model complexity by preferring models with fewer parameters. AIC and BIC have different desirable properties for model selection \cite{vriezeModelSelectionPsychological2012}, so rather than sticking to one, we have chosen to use both.

\begin{figure}[h!]
\centering
\includegraphics[width=1\textwidth]{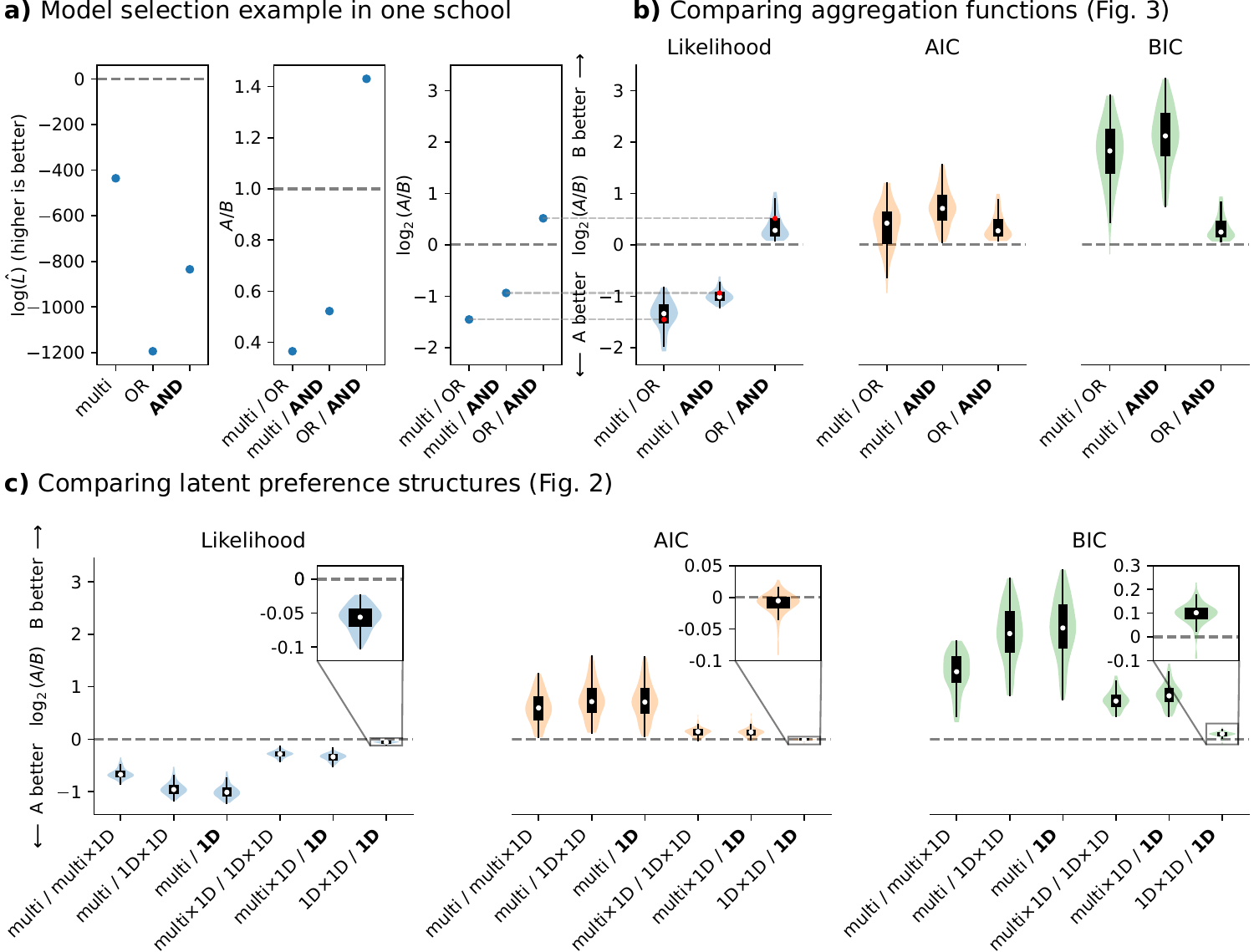}
\caption{\textbf{Testing preference structure and aggregation through model selection}. In panel \textbf{a} we illustrate how we compare models by computing model performance metrics (in this case, likelihood) in one particular school. We then plot the distributions of values obtained for the 70 schools considered. In panel \textbf{b}, we show the distributions of log ratios of maximum likelihood, AIC, and BIC for different aggregation functions. The aggregation functions at the top were applied to a preference structure of the type \textbf{1D} as represented in Fig.~\ref{fig:preference_dependencies}\textbf{d}. In panel \textbf{c}, we present the corresponding results for different latent preference structures. The preference structures at the bottom were aggregated using the \textbf{AND} aggregation function. Each violin plot and box plot combination represents the distribution of values obtained for all the analyzed schools of the AddHealth dataset. This result shows that the \textbf{AND} aggregation function with the \textbf{1D} preference structure has the overall best performance once the complexity of the model is taken into account through AIC and BIC.}
\label{fig:model_selection}
\end{figure}

To compare aggregation functions systematically, we have combined each of them with the simplest preference structure among those presented in Fig.~\ref{fig:preference_dependencies}: the 1D structure of panel \textbf{d}. We have also compared them to the full multidimensional model (panel \textbf{a}), which we use as a baseline. We have fit each friendship network of AddHealth independently, obtaining $\log (\hat{L})$, AIC, and BIC for all aggregation functions. Then, for each metric, we have performed pairwise comparisons of the aggregation functions and the multidimensional model by computing $\log_2 (\frac{A}{B})$, where $A =\log (\hat{L_a})$, AIC$_a$, or BIC$_a$, with $a$ being a combination of preference structure and aggregation function (and analogously for $B$). Since a lower absolute value of $|\log (\hat{L})|$, AIC, and BIC indicates a better fit, $\log_2 (\frac{A}{B})$ takes negative values if the combination in the numerator $A$ is better than the denominator $B$ and positive values otherwise. Panel \textbf{a} of Fig.~\ref{fig:model_selection} illustrates the likelihood computation for a single school. The violin plots in Fig.~\ref{fig:model_selection}\textbf{b} show the distribution of $\log_2 (\frac{A}{B})$ values for the 70 schools. For example, the first violin plot of the likelihood panel in Fig.~\ref{fig:model_selection}\textbf{b} corresponds to $\log_2 (\frac{\log(\hat{L}_{\text{multi}})}{\log(\hat{L}_{\text{1D}+\text{AND}})})$, and the third violin plot of the AIC panel corresponds to 
$\log_2 (\frac{AIC_{\text{1D}+\text{OR}}}{AIC_{\text{1D}+\text{AND}}})$.

We observe that the full multidimensional model always reaches higher values of log-likelihood than the 1D preferences aggregated with \textbf{AND} and \textbf{OR}, since the multidimensional model finds the exact parameters for which the likelihood function is maximized. By comparing the log-likelihood of the \textbf{AND} and the \textbf{OR} aggregation functions, we find that \textbf{AND} consistently outperforms the alternative. Since these two aggregation functions rely on the same number of parameters, the same pattern can be found by looking at AIC and BIC. Finally, we observe that once we consider the complexity of the model through AIC and BIC, both the \textbf{AND} and \textbf{OR} functions exceed the performance of the multidimensional model. This suggests that if we consider the combination of all attributes as a class on their own, as we do in the multidimensional model, we tend to overfit, and therefore simpler models are preferred. We have omitted the \textbf{MEAN} aggregation function from Fig.~\ref{fig:model_selection} because the results were almost identical to the \textbf{OR} aggregation function. In Methods section \ref{sec:or_and_mean_functions} we discuss the reason why these aggregation functions may lead to very similar preference estimations.

Since the \textbf{AND} aggregation function provides the best fit, we have combined it with each of the considered latent preference structures of Fig.~\ref{fig:preference_dependencies} to compare them. The results are shown in Fig.~\ref{fig:model_selection}\textbf{c}. For example, the fourth violin plot of the likelihood panel in Fig.~\ref{fig:model_selection}\textbf{c} corresponds to 
$\log_2 ( \frac{ \log(\hat{L}_{\text{multi}\times \text{1D} + \text{AND}}) }  { \log(\hat{L}_{\text{1D} \times \text{1D}+\text{AND}} )}) $.  As expected, complex models with more parameters outperform simpler models in terms of maximum likelihood. However, once we consider the model's complexity and penalize the number of parameters, simpler models are preferred. The two simplest models (1D and 1D$\times$1D) have comparable performance in terms of AIC; however, the simpler 1D model outperforms the more complex 1D$\times$1D model in terms of BIC. Furthermore, the latent preferences of the 1D model have a simpler and more meaningful interpretation, so we consider it the most likely to describe our mental representation of preferences.


Our comparison of aggregation functions neatly answers the question of how we integrate the information of our multidimensional identities when interacting with each other: we use the simplest mechanism by evaluating each one-dimensional preference independently and requiring all the partial evaluations to be successful in order to make the connection. This mechanism is conceptually equivalent to the additive aggregation mechanism proposed in the crossed categorization literature \cite{grigoryanCrossedCategorizationOutside2020}. To understand this conceptual equivalence, we need to bear in mind that in the typical experimental setting of crossed categorization studies, connection preferences take the form of scores assigned by a respondent to \emph{vignette} individuals (hypothetical persons with synthetic profiles), and the simplest way to aggregate the scores of all dimensions is to add them. However, in our stochastic network model, preferences take the form of probabilities, and the correct way to aggregate the probabilities of two events in the simplest way is to consider them independent and thus compute their product.

Current standard social network models such as Stochastic Actor-Oriented Models (SAOM) \cite{blockMultidimensionalHomophilyFriendship2014} or Exponential Random Graph Models (ERGM) \cite{wimmerRacialHomophilyERG2010} also operate under the assumption that preferences aggregate in a multiplicative way. In the formulation of both models, the existence probability of an edge $i\rightarrow j$ is $p_{ij} \propto e^{\boldsymbol{\beta} \mathbf{z}}$, where $\boldsymbol{\beta}$ is the vector of model parameters and $\mathbf{z}$ a vector of network statistics, including node group membership. One way researchers operationalize group connection biases is by including $v_d \times v_d$ dummy variables in the $\mathbf{z}$ vector for each dimension $d$, which are $1$ only if node $i \in r$ and node $j \in s$ of dimension $d$, and $0$ otherwise. For example, if two values are considered for the gender dimension, there would be 4 dummy indicator variables, one for each pair of groups ($\female\female, \female\male, \male\female, \male\male$). If we call these dummy variables $\mathbf{z}^d$, the terms of $p_{ij}$ related to group connection biases can be written as $p_{ij} \propto e^{\sum_d \boldsymbol{\beta}^d \mathbf{z}^d} = e^{\boldsymbol{\beta}^1 \mathbf{z}^1} e^{\boldsymbol{\beta}^2 \mathbf{z}^2} \cdots e^{\boldsymbol{\beta}^D \mathbf{z}^D}$. Notice that for each pair of nodes, only one of the terms of $\mathbf{z}^d$ will be nonzero. Therefore, for a particular pair of nodes belonging to specific multidimensional groups, $i\in\mathbf{r}, j\in \mathbf{s}$, we have $p_{ij} \propto e^{\beta^1_{r^1,s^1}} e^{\beta^2_{r^2,s^2}} \cdots e^{\beta^D_{r^D,s^D}} = \prod_d e^{\beta^d_{r^d,s^d}}$, which can be interpreted in an analogous way to Eq. \eqref{eq:all_dim}. With our systematic analysis, we show why the formulations of SAOM and ERGM were often successful in recovering underlying preferences, as they were using the aggregation function most likely to describe the data.

Combining these insights with our analysis of preference structures, we conclude that when forming ties, individuals represent and aggregate the information about their multidimensional identities in the simplest manner: by independently evaluating one-dimensional preferences that depend only on the dimension being assessed. For instance, the preferences of the multidimensional group (woman, Hispanic) are a combination of the preferences of women for people from other genders and Hispanic people for people of other ethnicities. This is a very significant implication because it means that, for example, a woman's preference for other women is independent of her ethnicity, or that this dependency is so weak that we can safely ignore it and still obtain an excellent description of the data.

Another broader consequence is that in contexts where attributes are uncorrelated, one-dimensional estimation of preferences may not result in particularly bad approximations of actual preferences. The reason is that in networks generated with preference structure 1D and aggregation function \textbf{AND}, the one-dimensional estimations recover the correct one-dimensional preferences if attributes are completely uncorrelated (see Fig.~\ref{fig:wrong_1d_preference} and Supp. Section \ref{sec:biased_1d_preferences}
).

\subsection{Inferring and analyzing latent preferences in real-world scenarios}
\label{sec:empirical_onedim_prefs}

Having identified the mechanisms through which multidimensional identities drive tie formation, we can now leverage this knowledge to uncover latent preferences in real systems. As we will show below, our framework helps us understand how social groups connect and, crucially, what are the salient attributes we use when interacting with each other.

Considering the findings of the previous section, we use the \textbf{AND} aggregation function with the 1D preference structure to obtain the latent preferences from the high school friendship networks of the AddHealth dataset. 
To make 1D preferences comparable across different networks and easier to interpret, we have normalized them by dividing each row of the one-dimensional preference matrices by the diagonal term; that is, the in-group preference of homophily $\frac{h_{r,s}}{h_{r,r}}$. We present the raw non-normalized preferences in Supp. Fig.~\ref{fig:onedim_preferences_unnorm}.

\begin{figure}[h!]
\centering
\includegraphics[width=0.9\textwidth]{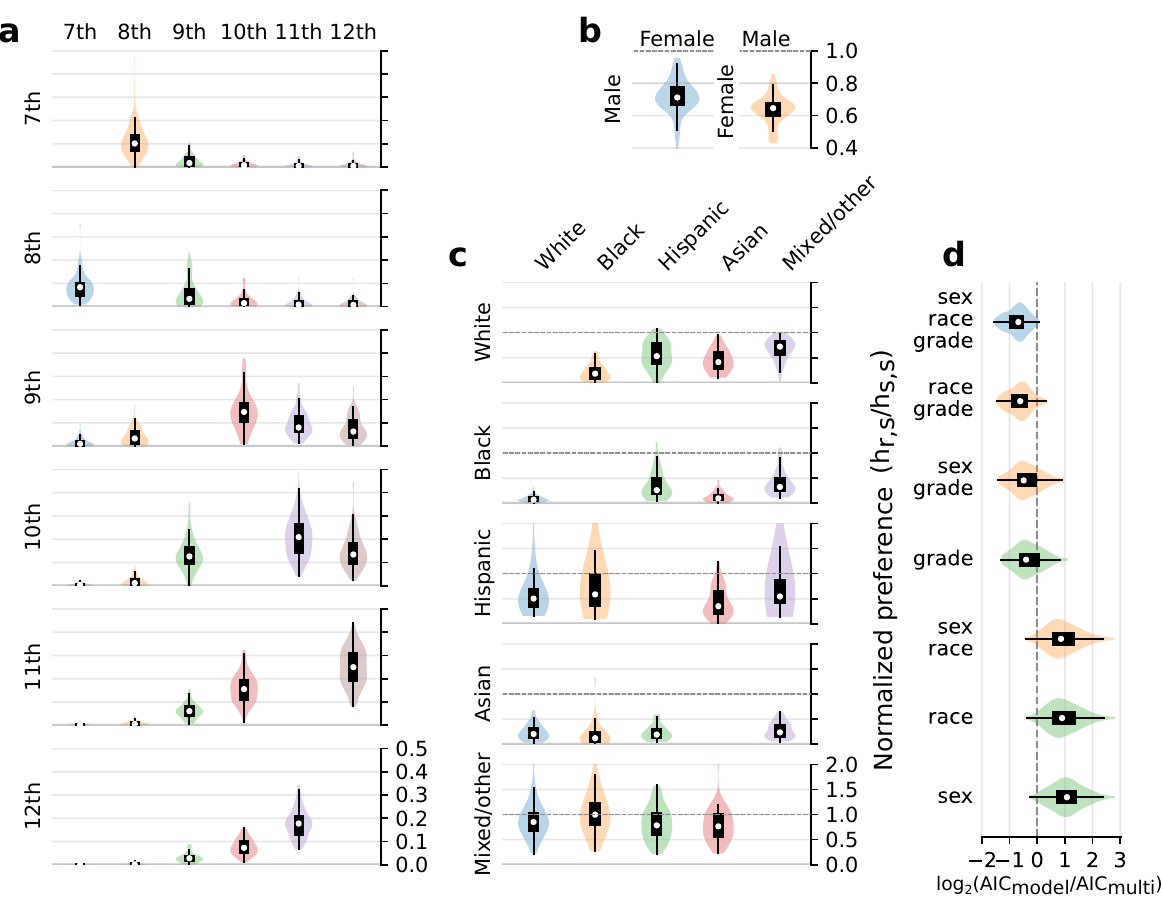}
\caption{\textbf{One-dimensional connection preferences in high school friendships (AddHealth)}. Panels \textbf{a}-\textbf{c} show the one-dimensional preferences obtained from fitting the model using grade, race, and gender as attributes. Preferences are obtained for each one-dimensional group in each school and then normalized by dividing by the in-group preference. Therefore, the normalized in-group preference is 1 by construction; we do not show it in this figure. Values lower than 1 in all the out-group preferences indicate a consistent homophily, while a value higher than 1 indicates heterophily for that particular group. Each violin plot and box plot combination represents the distribution of values obtained from fitting all the considered schools in the AddHealth dataset. Panel \textbf{d} shows $\log_2\left( \frac{AIC_a}{AIC_{multi}} \right)$ where $a$ is a 1D+\textbf{AND} model including only the dimensions shown in the y-axis. Lower values indicate better performance and the models are sorted from top to bottom by decreasing average performance. This result shows that grade is the most salient dimension, followed by race and sex.}
\label{fig:onedim_preferences}
\end{figure}

In Fig.~\ref{fig:onedim_preferences}, we show the distribution of preferences for the dimensions of grade, race, and gender. As in Fig.~\ref{fig:model_selection}, each violin plot represents the distribution of the 70 values obtained for all the considered schools. Since all preferences are normalized by in-group preference, a value higher than 1 indicates a higher tendency to connect to that group than to the in-group and vice versa. In the grade dimension, all preferences are lower than 1, indicating strong homophilic tendencies, as expected. In this dimension, preference is partially confounded with opportunity, as students naturally have more opportunities to interact with other schoolmates from their own grade. Nevertheless, the relative preferences show a remarkably regular and informative trend. Students prefer to connect with schoolmates of neighboring grades; however, this preference is not symmetric but aspirational. They prefer schoolmates of higher grades over those of lower grades. In the race dimension preferences show higher variation, sometimes reaching values above the in-group baseline, with students from different races presenting significantly different behaviors. Asian students are the most homophilic overall, while Mixed and Hispanic students are the most prone to cross-group connections. The latter are likely forming bridges across homogeneous groups, as some cross-group preferences are particularly low. For example, we can observe a mutual avoidance between Black and White students, which can be understood in the wider context of racial interactions in the United States. Finally, while all students are homophilic in terms of gender, boys show a slightly higher relative preference towards girls than the other way around. In Supp. Figs. \ref{fig:example_network_first}-\ref{fig:example_network_last} we show some representative examples of networks from the AddHealth dataset and their latent preferences.

Considering the values of relative preference, the dimension with the strongest connection biases is grade, which is expected. The second dimension would be race, although we observe large variability between groups (Asians are consistently homophilic while Hispanics show a wide variety of preferences). Sex is overall the most neutral dimension. This analysis provides a hierarchy of dimension salience, as a neutral dimension is, by definition, irrelevant to making connections \cite{ContemporarySocialPsychological2018}. We can use model selection to provide an even more robust assessment of this hierarchy by fitting a one-dimensional model and systematically including and excluding dimensions while measuring the change in model performance. The dimension that provides the best one-dimensional model is the most salient, followed by the one that provides the best two-dimensional model in combination with the first, and so on. We have performed this analysis by computing the logarithm of the ratio of the model's AIC to the multidimensional case. We report the results in Fig.~\ref{fig:onedim_preferences}\textbf{d}, verifying that grade is the most salient dimension, followed by race and sex.

Our network model does not explicitly include endogenous network processes, such as reciprocity and transitivity, but we can use random permutation techniques to estimate their influence and obtain p-values for the inferred preferences. We have implemented a methodology based on Multiple Regression Quadratic Assignment Procedure (MRQAP) \cite{dekkerSensitivityMRQAPTests2007}, which has been successfully applied in recent social network studies to control for endogenous network processes \cite{bodaEthnicDiversityFosters2023}. In essence, the method consists in computing a null distribution of preferences by 1) randomly swapping attribute vectors between nodes, so that we obtain a network with the same topology and same attribute distribution, but where the vectors of attributes belong to different nodes; and 2) computing the connection preferences of this new network. We repeat this procedure multiple times and compute a one-sided p-value by calculating the proportion of simulated preference values that are larger than the empirical one (if the empirical preference is lower than the simulated mean) or lower than the empirical preference (if the empirical is higher). In our analysis of the AddHealth dataset, we find that grade-based and sex-based preferences are always significant at a confidence level of $\alpha=0.01$, while race-based preferences are significant most of the time for all groups but the Mixed one (70\% of all normalized preferences for White, 83\% for Black, 77\% for Asian, 50\% for Hispanic, and 14\% for mixed). We can understand the lower overall significance for the mixed group by taking into account that they are the smallest group (see Supp. Fig. \ref{fig:group_sizes}), so their inferred preferences are the noisiest, and the fact that their group members are the most heterogeneous, both at the individual and at the collective level. For Hispanic, their low significance stems from the small \emph{effect size} of their preferences (after the mixed group, they are the closest to neutrality), and their relatively small sample size. We report the p-values for all schools and preferences in the results table provided as supplementary material. The coding scheme for the columns of the table is detailed in Supp. Section \ref{sec:coding_scheme}.


We have compared our results to earlier studies of the AddHealth database focused on race-based connection preferences. We have found that our inferred preferences for the race dimension are highly correlated ($r^2=0.91$) with those of Mouw and Entwisle \cite{mouwResidentialSegregationInterracial2006}, who used a logit model controlling for school diversity, sex, grade, geographical proximity, and parents' education and income. We provide further details about this comparison in the Methods section.

\section{Discussion}

{
Our identities are composed of multiple categories – race, gender, socioeconomic status – that shape our interactions. The question of how do we combine this multi-level information from our multidimensional identities to form social connections has long remained unanswered. Without answering this question, our attempts to understand our multi-level social networks are limited. 
In this work, we address this question by introducing a conceptual and mathematical framework to model social interactions between individuals with multidimensional identities. We have used it as a test bed to answer fundamental questions about multidimensional interactions, fitting the model with data from high school friendship networks. Firstly, we have identified the mental representations that structure our latent preferences towards other social groups. Secondly, we have revealed the aggregation mechanism we use to integrate information about our multidimensional identities when forming social ties. 
}

{
We have considered a collection of reasonable alternatives for the structures of preferences, and several aggregation mechanisms with clear and intuitive interpretations, from a mechanism where agents are tolerant and establish connections as soon as they manifest high preference in a single dimension, to a mechanism where agents are very selective and only form ties with those for whom they have high preference in all dimensions. We compared all the alternatives using Bayesian model selection and found that combining the simplest preference structure (one-dimensional preferences independent across dimensions) with a basic aggregation mechanism (independent evaluation of all dimensions) offers the most plausible description of the friendship network data. Our results provide a meaningful answer to the question raised by DiMaggio and Garip \cite{dimaggioNetworkEffectsSocial2012} about how would a female Hispanic college graduate choose her social ties: she would evaluate her preferences in all dimensions independently and form a tie if each individual evaluation is positive.
Our findings are in line with earlier studies of crossed categorization \cite{grigoryanCrossedCategorizationOutside2020} and connect with the rich body of cognitive science literature showing that we often follow a simplicity principle when facing cognitively demanding tasks \cite{feldmanSimplicityPrinciplePerception2016}.
}


{
We have further shown the applicability of the model to recover latent one-dimensional preferences in real-world multidimensional systems: high school friendship networks in the United States. To this end, we have derived a closed-form expression for the model's likelihood function. While Bayesian model selection validates the model quantitatively, this empirical analysis is a qualitative validation, as it demonstrates that the model provides meaningful interpretable results. For example, we have identified an aspirational linking tendency of students from lower grades for those of higher grades, widespread racial homophily, and the inter-racial preferences that may be expected in the context of North American schools, like a low mutual preference between black and white students. Moreover, the model provides natural operationalizations for the concept of dimension salience (and salience hierarchy), as conceptualized in social psychology \cite{ContemporarySocialPsychological2018}. We define dimensions as salient when their associated preferences depart substantially from neutrality and/or if they provide large likelihood improvements when included in the model.
}

{
With the increasing availability of longitudinal network datasets, a future iteration of our framework could explicitly incorporate dynamical network aspects, such as link formation and removal, or dynamical attributes to obtain more complete information about the evolution of connections. We could also incorporate endogenous tie formation mechanisms such as preferential attachment, triadic closure, or reciprocity into the inference process. As discussed earlier, the core element of the model – the mapping from multidimensional identities to connection probabilities – can be applied to any existing stochastic network model. Therefore, a promising avenue to incorporate these new elements consists in including the latent preference aggregation mechanism in novel network models that include endogenous processes like triadic closure \cite{peixotoDisentanglingHomophilyCommunity2022} and dynamical link evolution \cite{asikainenCumulativeEffectsTriadic2020}.
}

{
To conclude, our framework helps us reveal the fundamental preference structures and aggregation mechanisms people use when interacting with each other through multidimensional identities, addressing pressing but often neglected questions in social psychology. We have explicitly demonstrated the hazards of ignoring multidimensionality if attributes are correlated and we have proposed natural operationalizations for the concept of dimension salience.
Our model of latent preferences and their aggregation can be applied to a variety of network models, allowing researchers to better examine the effects of multidimensional identities on dynamical processes such as diffusion of information, formation of norms, and dissemination of culture. We can readily use this method to measure the innate preferences people use to form ties in real-world social networks, allowing us to better address far-reaching social issues like the emergence of multidimensional inequalities in networks.
}

\section{Methods}
\subsection{Mapping attribute vectors to numerical indices}


\NewDocumentCommand{\dslash}{s}{%
  \IfBooleanTF{#1}
    {\big/\mkern-7mu\big/}
    {/\mkern-4mu/}%
}

In this section, we explain the mapping we use to switch between vectors of attributes $\mathbf{s}$ and integer indices when working with matrices such as $H_{\mathbf{r},\mathbf{s}}$. In this section, instead of starting the indices from $1$, we start from $0$, so that for each attribute $d$, $s^d\in\{0,1,\dots,v_d-1\}$. 
As a preliminary step, we need to fix the indices of the dimensions; for instance, in the example illustrated in the main document with gender and nationality as the relevant attributes, we fix gender to be $d=1$ and nationality to be $d=2$. Then, within each dimension, we need to fix the mapping between categories and numerical indices; for instance, $\{\female \rightarrow 0, \male\rightarrow 1 \}$ for the gender dimension.
We now construct a mapping that links the vector of indices $\mathbf{s}=(s^1,\dots,s^D)$ to a unique integer $s^*$ in the set $\{0,1,\dots,\mathcal{G}-1\}$:
\begin{align*}
    s^* &= s^1 + v_1\cdot s^2 + v_1 v_2\cdot s^3 + \dots + \prod_{d=1}^{D-1}v_d \cdot s^D = \sum_{l=1}^D s^l \prod_{d=1}^{l-1}v_d
\end{align*}
The inverse operation, from integer index $s^*$ to vector $\mathbf{s}$, can be performed using the following equations:
\[ \begin{cases}
    s^1 = mod(s^*,v_1) \\
    s^2 = mod(s^*\dslash v_1, v_2) \\
    s^3 = mod(s^*\dslash v_1v_2, v_3)
    \\
    \vdots \\
    s^D = mod(s^*\dslash \prod_{d=1}^{D-1}v_d, v_D)
\end{cases}\]
where $mod$ is the modulus operator and $\dslash$ the truncating integer division.

\subsection{Model identifiability and number of free parameters}
\label{sec:num_parameters_identifiability}

The lower-dimensional matrices $\mathbf{h}$ aggregated through the aggregation functions presented in Section \ref{sec:aggregation_functions} are unidentifiable by construction up to a number of constants that coincide with the number of pairs of latent lower-dimensional preference matrices. To see this, consider a two-dimensional system with two preference matrices $\mathbf{h}^1$ and $\mathbf{h}^2$. Given this pair of matrices, we can obtain another pair $\mathbf{h}^{1*},\mathbf{h}^{2*}$ that lead to the same multidimensional preference matrix $\mathbf{H}$ using each of the aggregation functions:

\begin{align} 
\textbf{AND} \longrightarrow & H_{(r^1,r^2),(s^1,s^2)} = h^1_{r^1,s^1} h^2_{r^2,s^2}\\
&\mathbf{h}^{1*}=C \mathbf{h}^1  ;\quad  \mathbf{h}^{2*} = \frac{1}{C} \mathbf{h}^2 \\
\textbf{MEAN} \longrightarrow & H_{(r^1,r^2),(s^1,s^2)} = \frac{1}{w^1 + w^2}(w^1 h^1_{r^1,s^1} + w^2 h^2_{r^2,s^2}) \\
& \mathbf{h}^{1*} = \mathbf{h}^1 + C ;\quad  \mathbf{h}^{2*} = \mathbf{h}^2 - \frac{w^1}{w^2} C \\
\textbf{OR} \longrightarrow & H_{(r^1,r^2),(s^1,s^2)} = 1-(1-h^1_{r^1,s^1})(1- h^2_{r^2,s^2}) = h^1_{r^1,s^1} + h^2_{r^2,s^2} - h^1_{r^1,s^1} h^2_{r^2,s^2} \\
&\mathbf{h}^{1*} = C \mathbf{h}^1 + (1-C) ;\quad  \mathbf{h}^{2*} = \frac{1}{C} \mathbf{h}^2 + (1-\frac{1}{C}) \\
\end{align}

Here, $C$ is a constant that can only take values such that the elements of the new matrices $h^{d*}_{r^d,s^d} \in [0,1]$. Despite this constraint, it is in general possible to find some $C \neq 1$ leading to several non-trivial pairs of matrices producing the same $\mathbf{H}$. Although in the example there are only two dimensions, these relationships can be established between any pair of matrices in a system with an arbitrary number of dimensions. 

The number of lower-dimensional preference matrices depends on the structure of preferences, as illustrated in Fig.~\ref{fig:preference_dependencies} of the main document. The structures $\text{multi} \times 1\text{D}$ (panel \textbf{b} of Fig.~\ref{fig:preference_dependencies}) and $1\text{D}$ (panel \textbf{d}) have $D$ lower-dimensional matrices, amounting to $\binom{D}{2}$ pairs. The structure $1\text{D} \times 1\text{D}$ (panel \textbf{c}) has $D^2$ lower-dimensional matrices, so there are $\binom{D^2}{2}$ pairs. As a result of these pairwise dependencies between the latent lower-dimensional preference matrices, the number of free parameters for each preference structure is lower than the number of elements in the preference matrices, as shown in Table \ref{tab:free_params}.

\begin{table}
\caption{Number of free parameters of each preference structure.}
    \centering
    \begin{tabular}{p{0.23\textwidth}|p{0.23\textwidth}|p{0.23\textwidth}|p{0.23\textwidth}}
         Preference structure & Parameters of the lower-dimensional matrices & Number of matrices & Number of free parameters\\ \hline
         multi & $(\prod_d v_d )^2$ & 1& $(\prod_d v_d )^2$\\
         $\text{multi} \times 1\text{D}$& $\prod_d v_d  \sum_d{v_d}$ & $D$ &$\prod_d v_d  \sum_d{v_d} - \binom{D}{2}$\\
         $1\text{D} \times 1\text{D}$& $(\sum_d{v_d} )^2$& $D^2$ & $( \sum_d{v_d} )^2  - \binom{D^2}{2}$\\
         1D& $\sum_d {v_d}^2$ & $D$ & $\sum_d {v_d}^2  - \binom{D}{2}$\\
    \end{tabular}
    \label{tab:free_params}
\end{table}

To fully and unambiguously determine the lower-dimensional preference matrices, we need to impose a number of constraints that matches the reduction in the number of free parameters; that is, the number of pairs of matrices. One possibility is to require that (some of) the $\textbf{h}$ matrices be normalized so that their elements sum up to $1$, thus imposing up to $D$ constraints, one for each matrix. Focusing on the preference structure $1\text{D}$, which is the one for which we find the most empirical support, this normalization would be enough for $D=2$ and $3$ dimensions, because $\binom{2}{2}=1$ and $\binom{3}{2}=3$. However, when the number of pairs of matrices is higher than $D$, these constraints are not enough to unambiguously determine all the $\textbf{h}$. For example, with $D=4$ dimensions, there are $\binom{4}{2}=6$ pairs of matrices. To circumvent this unidentifiability issue in a higher-dimensional setting, we can normalize $\mathbf{h}$ by dividing each row by its diagonal term (in-group preference or homophily), so that the preference matrices are unique and retain the relevant information about relative cross-group preferences. This normalization works as long as $\sum_d v_d > \binom{D}{2}$; for example, if $v_1=\cdots=v_D = 2$, it would work up to $D=5$. Although the data we consider only includes up to $3$ dimensions, we use this row-wise normalization because it enables the comparison across groups and networks with different densities, number of nodes, and number of interacting groups.

In practice, we fit the unconstrained model and normalize the matrices afterward. We find that the results are fully consistent across different fitting methods and parameter initializations. After all, even if there are several parameter combinations that optimize the likelihood, all are equivalent once adequate constraints are imposed.


\subsection{Similarity between the OR and the MEAN aggregation functions}
\label{sec:or_and_mean_functions}

In this section, we will analyze the similarities between then \textbf{OR} and \textbf{MEAN} aggregation functions. We will use the 1D preference structure of Fig.~\ref{fig:preference_dependencies}\textbf{d}.
Let us fix the two interacting multidimensional groups $\mathbf{r}$ and $\mathbf{s}$, so that we need to compute element $H_{\mathbf{r},\mathbf{s}}$ of the multidimensional preference matrix, which depends on the vector of latent preferences $\mathbf{h}=(h_{r^1,s^1}^1,h_{r^2,s^2}^2,\dots,h_{r^D,s^D}^D)$. To simplify the notation, let us call element $H_{\mathbf{r},\mathbf{s}}$ simply $H$, and the elements of the latent preference vector, $h^d$. With these considerations, the definitions of the \textbf{OR} and \textbf{MEAN} functions are:
\[H_{MEAN} = \frac{1}{D} \sum_{d=1}^D h^d \qquad \qquad \qquad\qquad\qquad H_{OR} = 1 - \prod_{d=1}^D (1-h^d)  \]
We can notice now that if we expand the product inside the OR function, we obtain:
\begin{align*}
    H_{OR} &= 1 - \prod_{d=1}^D (1-h^d) = 1 - \left( 1 - \sum_{d=1}^D h^d + \sum_{1 \leq i < j \leq D} h^i h^j - \sum_{1 \leq i < j < k \leq D} h^i h^j h^k + \cdots + (-1)^D \prod_{d=1}^D h^d \right) \\
    &= \sum_{d=1}^D h^d - \sum_{1 \leq i < j \leq D} h^i h^j + \sum_{1 \leq i < j < k \leq D} h^i h^j h^k - \cdots + (-1)^{D+1} \prod_{d=1}^D h^d 
\end{align*} 
For small values of $h^d$, the higher-order terms are negligible, and therefore the first-order approximation of the OR function is equivalent to $\sum_{d=1}^D h^d$, which is proportional to the MEAN aggregation function.

Finally, we show that the proportionality between $\sum_{d=1}^D h^d$ and the MEAN function is actually an equivalence in the model. Let us suppose that we find parameters $\bar h^d$ that maximize the likelihood for the MEAN function, and such that $0\le \bar h^d \le 1$. Then, parameters $\tilde h^d = \frac{1}{D} \bar h^d$ will also maximize the likelihood when we optimize the model for the aggregation function $\sum_{d=1}^D h^d$. 

This leads to the conclusion that, when $h^d$ is small, the MEAN and the OR aggregation functions will learn very similar optimal parameters and their corresponding maximum likelihood can be expected to be comparable. Since they also use the same number of parameters, AIC and BIC also take similar values.


\subsection{Comparison with previous studies}

We have compared our results to earlier studies using Table 4 of \cite{zengPreferenceOpportunityChoice2008}, which summarizes the race-based connection preferences of three different studies of the AddHelath dataset: those by Quillian and Campbell \cite{quillianRaceClassSchool}, Mouw and Entwisle \cite{mouwResidentialSegregationInterracial2006}, and Zeng and Xie \cite{zengPreferenceOpportunityChoice2008}. The preferences shown in \cite{zengPreferenceOpportunityChoice2008} are expressed in cross-group connection odds ratios normalized by the in-group odds ratio, so we have performed the same computation starting from the original connection probabilities inferred by the model (instead of the normalized version of connection probabilities shown in Fig.~\ref{fig:onedim_preferences}). Thus, we have first computed $\text{ODDS}=\frac{p}{1-p}$ from each one-dimensional connection probability, and then we have normalized by in-group $\text{ODDS}$. To perform the comparison to the other studies we have computed the mean of the $\text{ODDS}$ for all networks weighted by the product of the sizes of the two interacting groups. We have found that our results are highly correlated with those obtained by Mouw and Entwisle \cite{mouwResidentialSegregationInterracial2006} ($r^2=0.91$), with lower correlations with Quillian and Campbell's results ($r^2 = 0.44$) and Zeng and Xie ($r^2 = 0.20$). While the study by Mouw and Entwisle uses a rudimentary logit model, it controls for school diversity, sex, grade, geographical proximity, and parents' education and income.

\section*{Acknowledgments}

This research work was funded by the European Union under the Horizon Europe MAMMOth project, Grant Agreement ID: 101070285, and by the Austrian Research Promotion Agency (FFG) under project No. 873927.

\bibliography{bibliography/zotero}
\bibliographystyle{splncs04}


\newpage
\begin{center}
\textbf{\LARGE Supplementary Information}
\end{center}
\setcounter{equation}{0}
\setcounter{figure}{0}
\setcounter{table}{0}
\setcounter{page}{1}
\setcounter{section}{0}
\makeatletter
\renewcommand{\theequation}{S\arabic{equation}}
\renewcommand{\thesection}{S\arabic{section}}
\renewcommand{\thefigure}{S\arabic{figure}}
\renewcommand{\thetable}{S\arabic{table}}

\section{Data description}
\label{sec:appendix_data}

The AddHealth dataset was built through a social survey carried out in 84 communities in 1994-95. Some of these communities had one high school and others two, usually split into junior high and high school. Each student was given a paper-and-pencil questionnaire and a copy of a roster listing every student in the school (or schools, if there were two). The name generator asked about five male and five female friends separately. The question was, "List your closest (male/female) friends. List your best (male/female) friend first, then your next best friend, and so on. (girls/boys) may include (boys/girls) who are friends and (boy/girl) friends." For each friend named, the student was asked to check off whether he/she participated in any of five activities with the friend: going to her house, meeting after school to hang out, spending time together during the weekend, talking about a problem in the last seven days, and talking on the phone in the last seven days. For our analyses, we have used all the friendship nominations regardless of the number or type of activities shared by the students.

One potential issue given that the question asks about male and female friends separately is that gender preference may be underestimated. However, students rarely provide a full list of 10 friends (out of all respondents, only 3\% do). Furthermore, we have found gender homophily and a clear asymmetry in gender preferences, so despite the question's wording, we are able to capture gender-based connection biases.

To prepare the data for the analysis, we removed all the nodes with unreported attributes. We also removed one-dimensional groups with fewer than 20 individuals to reduce the noise of the results. Our results are very robust to this filtering, as the inferred preferences for the groups that were not removed remain almost unchanged. Finally, we only analyzed schools with more than 100 students remaining in the network after filtering and with more than one gender and race category.  As a result, 41880 students from 70 communities were included in our study. We have analyzed each of the 70 communities independently.

\section{Biased one-dimensional preferences in multidimensional systems}
\label{sec:biased_1d_preferences}

In this section, we show how naively inferring one-dimensional preferences in multidimensional systems by applying the 1D MLE of Eq.~\eqref{eq:1d_mle}  (in the main document) one dimension at a time leads to erroneous results. We demonstrate this by deriving a mean-field analytical expression for the value of the 1D MLE in one of the two dimensions of a two-dimensional system. 

Let us assume that we have a two-dimensional system with two groups per dimension, so that the attribute vectors can take values $\mathbf{s} \in \{(1,1),(1,2),(2,1),(2,2)\}$. The nodes have 1D latent preferences as in Fig.~\ref{fig:preference_dependencies}\textbf{d} of the main document, encoded in two matrices $h^1_{r^1,s^1}, h^2_{r^2,s^2}$.

Let us change the notation for the 1D MLE of Eq.~\eqref{eq:1d_mle}, called $\hat{h}_{r,s}$ in the main document, to avoid confusion with the \emph{true} latent preferences $h^1_{r^1,s^1}, h^2_{r^2,s^2}$. In the remaining of this section, we will call it $\eta_{r,s}$ instead of $\hat{h}_{r,s}$. Since we can compute one $\eta_{r,s}$ matrix for each dimension, we will label them analogously to the $h$ matrices: $\eta^d_{r^d,s^d}$. Let us call $e^d_{r^d,s^d}$ the number of inter-group links from one-dimensional group $r^d$ to group $s^d$. With these considerations, we rewrite the 1D MLE of Eq.~\eqref{eq:1d_mle} as:

\begin{equation}
\label{eq:app_1d_mle}
    \eta^d_{r^d,s^d} = \frac{e^d_{^dr,s^d}}{N^d_{r^d}(N^d_{s^d}  - \updelta_{r^d,s^d})} \approx \frac{e^d_{r^d,s^d}}{N^d_{r^d} N^d_{s^d}}
\end{equation}

Where $N^d_{s^d}$ is the number of nodes belonging to one-dimensional group $s^d$. If we define $E_{\mathbf{r}, \mathbf{s}}$ as the number of links from multidimensional group $\mathbf{r}$ to multidimensional group $\mathbf{s}$, we can compute $e^d_{r^d,s^d}$ as:

\begin{equation}
    e^d_{r^d,s^d} = \sum_{\substack{\rho^d=r^d\\\sigma^d=s^d}} E_{\boldsymbol{\rho}, \boldsymbol{\sigma}}
\end{equation}

Here, the sum is over all multidimensional vectors $\boldsymbol{\rho}$ and $\boldsymbol{\sigma}$ whose $d$ component is respectively $r^d$ and $s^d$. Specifically, for a two-dimensional system with two groups per dimension, we have:

\begin{align}
    e^1_{r^1,s^1} =& E_{(r^1,1),(s^1,1)} + E_{(r^1,1),(s^1,2)} + E_{(r^1,2),(s^1,1)} + E_{(r^1,2),(s^1,2)}
    \\
    e^2_{r^2,s^2} =& E_{(1,r^2),(1,s^2)} + E_{(1,r^2),(2,s^2)} + E_{(2,r^2),(1,s^2)} + E_{(2,r^2),(2,s^2)}
    \label{eq:app_number_onedim_links_d2}
\end{align}

Using the aggregation function \textbf{AND}, the probability of connection across multidimensional groups is $H_{\mathbf{r}, \mathbf{s}} = h^1_{r^1,s^1} h^2_{r^2,s^2}$, and the expected number of links across two multidimensional groups is:

\begin{equation}
\label{eq:app_expected_number_multid_links}
    E_{\mathbf{r}, \mathbf{s}} = H_{\mathbf{r}, \mathbf{s}} N F_{\mathbf{r}} N F_{\mathbf{s}}
    = h^1_{r^1,s^1} h^2_{r^2,s^2} F_{\mathbf{r}} F_{\mathbf{s}} N^2
\end{equation}

Where $N$ is the total number of nodes. Now, let us compute $\eta^2_{r^2,s^2}$ by plugging Eq.~\eqref{eq:app_expected_number_multid_links} into Eq.~\eqref{eq:app_number_onedim_links_d2} and the result into Eq.~\eqref{eq:app_1d_mle}:

\begin{equation}
\begin{split}
    \eta^2_{r^2,s^2} \approx &\frac{N^2 \left[ 
    h^1_{1,1} h^2_{r^2,s^2} F_{(1,r^2)} F_{(1,s^2)} +
    h^1_{1,2} h^2_{r^2,s^2} F_{(1,r^2)} F_{(2,s^2)} +
    h^1_{2,1} h^2_{r^2,s^2} F_{(2,r^2)} F_{(1,s^2)} +
    h^1_{2,2} h^2_{r^2,s^2} F_{(2,r^2)} F_{(2,s^2)}
    \right] }{N^2_{r^2} N^2_{s^2}} 
    \\
    = &\frac{N^2 h^2_{r^2,s^2} \left[ 
    h^1_{1,1} F_{(1,r^2)} F_{(1,s^2)} +
    h^1_{1,2} F_{(1,r^2)} F_{(2,s^2)} +
    h^1_{2,1} F_{(2,r^2)} F_{(1,s^2)} +
    h^1_{2,2} F_{(2,r^2)} F_{(2,s^2)}
    \right] }{N^2_{r^2} N^2_{s^2}}
\end{split}
\end{equation}

From this result, it is already clear that in general $\eta^2_{r^2,s^2} \neq h^2_{r^2,s^2}$, so we should not use the 1D MLE to estimate preferences one dimension at a time in a multidimensional system.

To match the example discussed in the main document, we now assume that the population distribution is balanced dimension-wise. Therefore, the elements of the population fraction tensor fulfill the following constraints:

$$F_{1,1}+F_{1,2} = F_{2,1}+F_{2,2} = F_{1,1}+F_{2,1} = F_{1,2}+F_{2,2} = 0.5$$

This choice constraints the population distribution up to one free parameter, which controls correlation. To fully determine $F$, we tune the correlation parameter $\kappa$ to interpolate between the population distribution with maximum values in the diagonal and the one with maximum values in the anti-diagonal:

\begin{equation}
    \begin{NiceMatrixBlock}[auto-columns-width]
    F =  \kappa
        \begin{bNiceMatrix}
        0.5 & \quad 0 \\
        0 & \quad 0.5
        \end{bNiceMatrix}
    + (1-\kappa)
        \begin{bNiceMatrix}
        0 &\quad 0.5 \\
        0.5 &\quad 0
        \end{bNiceMatrix}
        \end{NiceMatrixBlock}
        =
        0.5
        \begin{bNiceMatrix}
        \kappa & \quad (1-\kappa) \\
        (1-\kappa) & \quad \kappa
        \end{bNiceMatrix}
\end{equation}



With these considerations, we can obtain closed-form expressions for the values of $\eta^2_{r^2,s^2}$:

\begin{align}
    \eta^2_{1,1} \approx& h^2_{1,1} [ 
    h^1_{1,1} \kappa^2 &+&
    h^1_{1,2} \kappa (1-\kappa) &+&
    h^1_{2,1} \kappa (1-\kappa) &+&
    h^1_{2,2} (1-\kappa)^2
    &]
    \\
    \eta^2_{1,2} \approx& h^2_{1,2} [ 
    h^1_{1,1} \kappa (1-\kappa) &+&
    h^1_{1,2} \kappa^2 &+&
    h^1_{2,1} (1-\kappa)^2 &+&
    h^1_{2,2} \kappa (1-\kappa)
    &]
    \\
    \eta^2_{2,1} \approx& h^2_{2,1} [
    h^1_{1,1} \kappa (1-\kappa) &+&
    h^1_{1,2} (1-\kappa)^2 &+&
    h^1_{2,1} \kappa^2 &+&
    h^1_{2,2} (1-\kappa)^2
    &]
    \\
    \eta^2_{2,2} \approx& h^2_{2,2} [ 
    h^1_{1,1} (1-\kappa)^2 &+&
    h^1_{1,2} \kappa (1-\kappa) &+&
    h^1_{2,1} \kappa (1-\kappa) &+&
    h^1_{2,2} \kappa^2
    &] 
\end{align}

The key insight that we obtain from these equations is that $\eta^2_{r^2,s^2}$ depends not only on the preferences in dimension $d=1$, but also on the population distribution. Furthermore, for a fixed population distribution (fixed $\kappa$), since the factors that multiply each of the matrix elements $h^1_{r^1,s^1}$ are different for each $\eta^2_{r^2,s^2}$, we can not simply rescale matrix $\eta^2_{r^2,s^2}$ to obtain $h^2_{r^2,s^2}$. This is nevertheless possible if attributes are uncorrelated, as in that case $F_{s^1,s^2} = f^1_{s^1}f^2_{s^2}$, and the expression for $\eta^2_{r^2,s^2}$ becomes:

\begin{equation}
    \eta^2_{r^2,s^2} = h^2_{r^2,s^2} [ 
    h^1_{1,1} f^1_1 f^1_1+
    h^1_{1,2} f^1_1 f^1_2+ 
    h^1_{2,1} f^1_2 f^1_1+
    h^1_{2,2} f^1_2 f^1_2 
    ] \propto h^2_{r^2,s^2} 
\end{equation}

Independently of the one-dimensional marginal population distributions.

\newpage
\section{Distribution of group sizes}

\begin{figure}[h!]
\centering
\includegraphics[width=\textwidth]{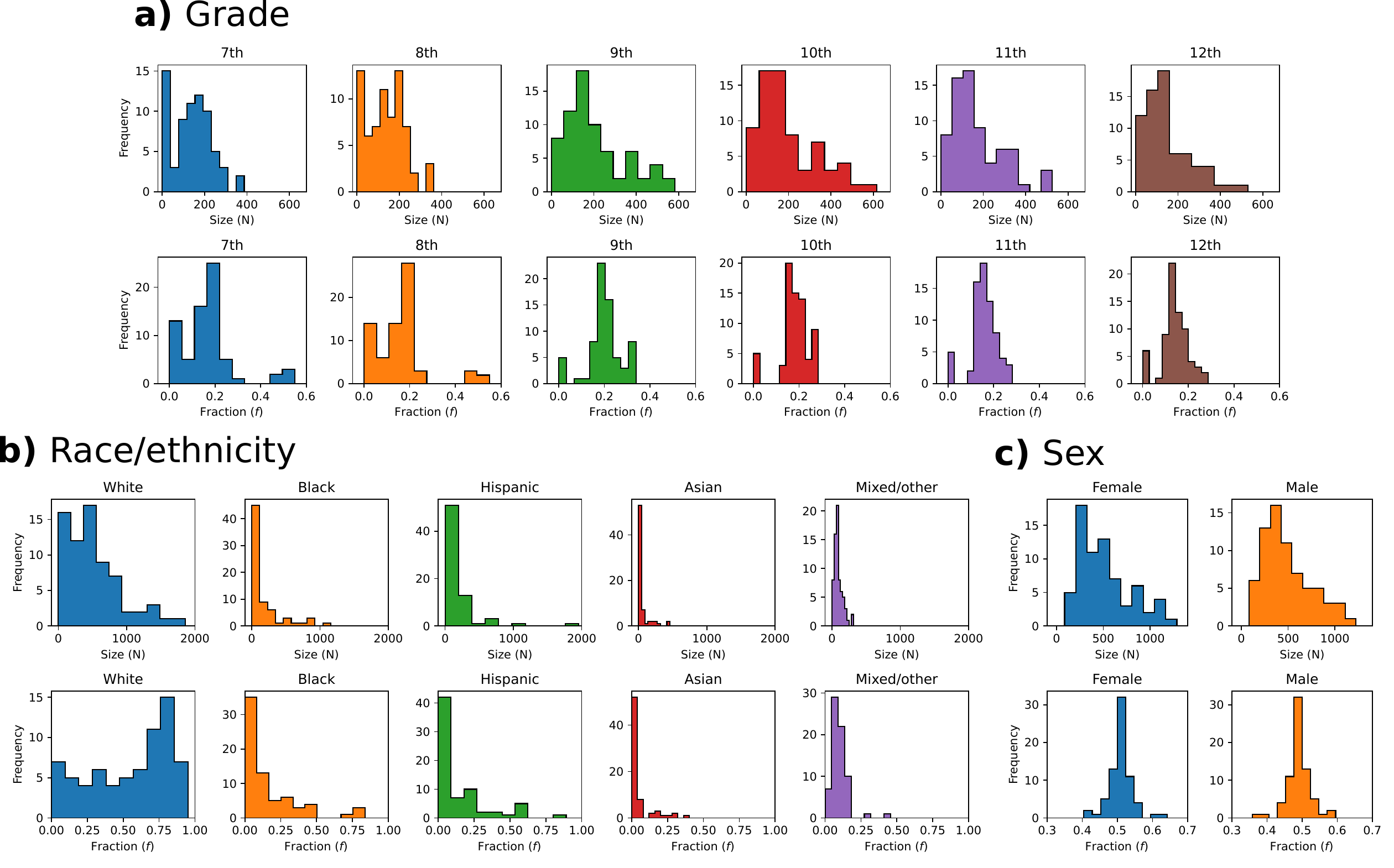}
\caption{\textbf{Distribution of group sizes in the AddHealth networks}. In each panel the top plots show the absolute size distribution in numbers of nodes and the bottom plots the relative size distributions as a fraction of the total population in each network.}
\label{fig:group_sizes}
\end{figure}

\newpage
\section{Unnormalized latent preferences}

\begin{figure}[h!]
\centering
\includegraphics[width=0.9\textwidth]{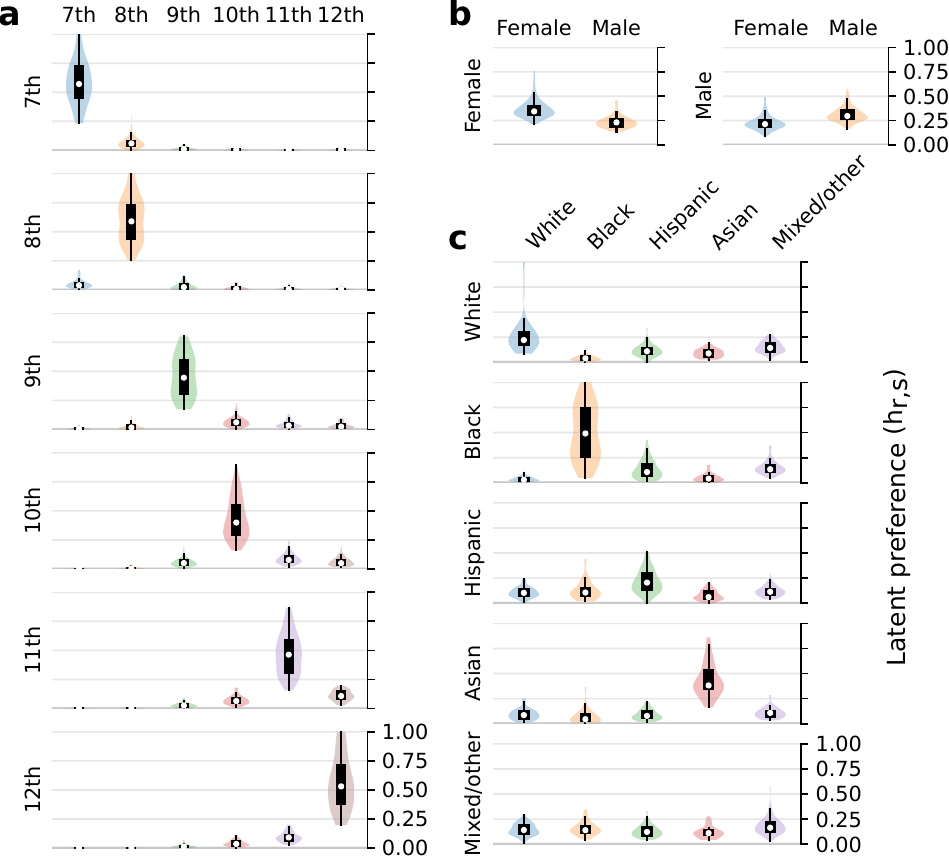}
\caption{\textbf{One-dimensional connection preferences in high school friendships (AddHealth)}. Non-normalized version of the preferences shown in Fig.~\ref{fig:onedim_preferences} in the main document. Each violin plot and box plot combination represents the distribution of values obtained from fitting all the considered schools in the AddHealth dataset.}
\label{fig:onedim_preferences_unnorm}
\end{figure}

In Fig.~\ref{fig:onedim_preferences_unnorm}, we show the non-normalized version of the $h^d_{r^d,s^d}$ matrices presented in Fig.~\ref{fig:onedim_preferences} in the main document.

\newpage
\section{Coding scheme for the supplementary results table}
\label{sec:coding_scheme}

We share all the final results obtained from the computations performed in this paper in a CSV file. In this file, each row corresponds to a school from the AddHealth dataset \cite{peixotoAdd_healthAdolescentHealth}. The table columns' name structure is as follows:

\begin{outline}
\1 School identifier: \texttt{school}
\1 Population size columns:
    \2 One-dimensional populations: \texttt{N\_[attribute]} where \texttt{[attribute]} is the name of the one-dimensional group (e.g. \texttt{N\_Asian}).
    \2 Multidimensional populations: \texttt{N\_[attr1]$|$[attr2]$|$[attr3]} where \texttt{([attr1], [attr2], [attr3])} is the multidimensional attribute vector (e.g. \texttt{N\_9th$|$Asian$|$Male}).
\1 Multidimensional preference columns: They all have the structure 

\texttt{H\_[aggregation function]\_[preference structure]\_[source multidim. group]-[target multidim. group]}
    \2 \texttt{$[$aggregation function$]$} can be \{\texttt{and,or,mean}\}.
    \2 \texttt{$[$preference structure$]$} can be:
            \3 \texttt{multi-full}: fully multidimensional, as in Fig.~\ref{fig:preference_dependencies}a (multi). This structure has no preceding \texttt{[aggregation function]} label, as it is unnecessary. 
            \3 \texttt{multi-1d}: as in Fig.~\ref{fig:preference_dependencies}b (multi$\times$1D).
            \3 \texttt{1d-full}: as in Fig.~\ref{fig:preference_dependencies}c (1D$\times$1D).
            \3 \texttt{1d-simple}: as in Fig.~\ref{fig:preference_dependencies}d (1D).
    \2 \texttt{$[$source multidim. group$]$} and \texttt{$[$target multidim. group$]$} are labeled as for the multidimensional populations \texttt{$[$attr1$]$$|$$[$attr2$]$$|$$[$attr3$]$} (e.g. 9th$|$Asian$|$Male).
\1 One-dimensional preference columns:
    \2 The column can start with \texttt{h\_} or with \texttt{h\_norm}:
        \3 \texttt{h\_} is for the raw inferred preference values.
        \3 \texttt{h\_norm} is for the in-group-normalized values, as explained in the paper.
    \2 For all preference structures but 1D, they have the structure 
    
    \texttt{h\_[aggregation function]\_[preference\_structure]\_[source group]-[target group]}
    \2 For preference structure 1D, they have the structure 
    
    \texttt{h\_[aggregation function]\_[preference\_structure]\_[dimension]\_[source group]-[target group]}, where \texttt{[dimension]} can be \{\texttt{grade,race,sex}\}.
\1 MRQAP columns. These columns have the structure 

{\scriptsize
\texttt{MRQAP\_[quantity]\_[h / h\_norm]\_[aggregation function]\_[preference structure]\_[dimension]\_[source group]-[target group]}}. Here, \texttt{[quantity]} can be:
    \2 \texttt{pval1s}: numerically computed p-value from 100 MRQAP simulations. Proportion of MRQAP randomizations where preference is less than the empirical.
    \2 \texttt{pval2s}: numerically computed p-value from 100 MRQAP simulations. Proportion of MRQAP randomizations where preference is either less or more than the empirical, whichever is lower. This is not the two-sided, which would be 2 times that value if the distribution is symmetric.
    \2 \texttt{av}: average value of one-dimensional preference h from 100 simulations.
    \2 \texttt{std}: standard deviation of one-dimensional preference h from 100 simulations.
\1 Model performance metrics columns. The structure is \texttt{[metric]\_[aggregation function]\_[preference structure]}, where \texttt{[metric]} can be \{\texttt{L,AIC,BIC}\}.
\end{outline}

\newpage
\section{Examples of AddHealth networks and their latent preferences}

\begin{figure}[h!]
\centering
\includegraphics[width=0.85\textwidth]{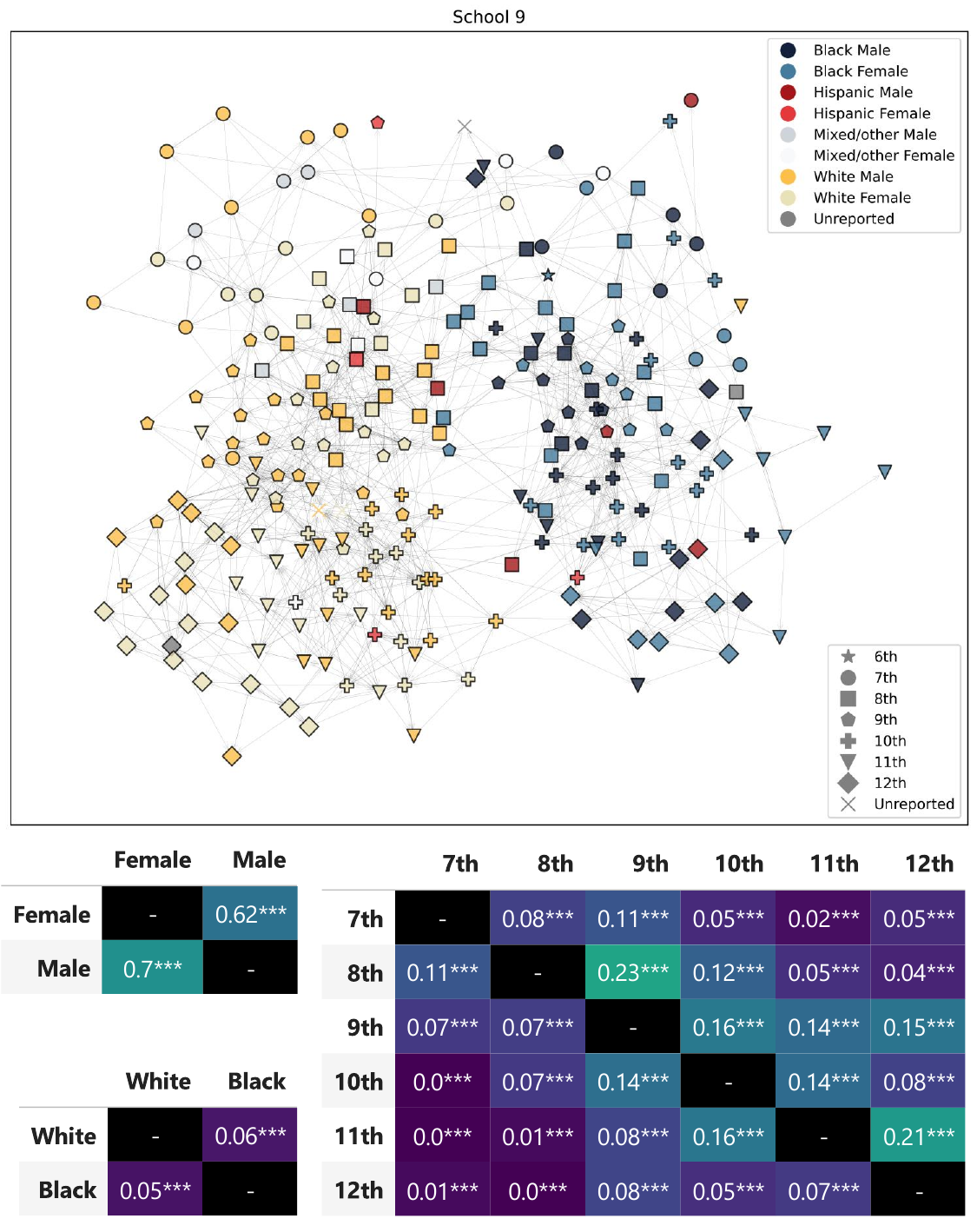}
\caption{Representative example of one of the AddHealth networks and its in-group-normalized latent connection preferences inferred for a 1D+\textbf{AND} model. Asterisks indicate the statistical significance computed using the MRQAP method: $\text{p-value}<0.05^*, 0.01^{**}, 0.001^{***}$.}
\label{fig:example_network_first}
\end{figure}

\begin{figure}[h!]
\centering
\includegraphics[width=0.95\textwidth]{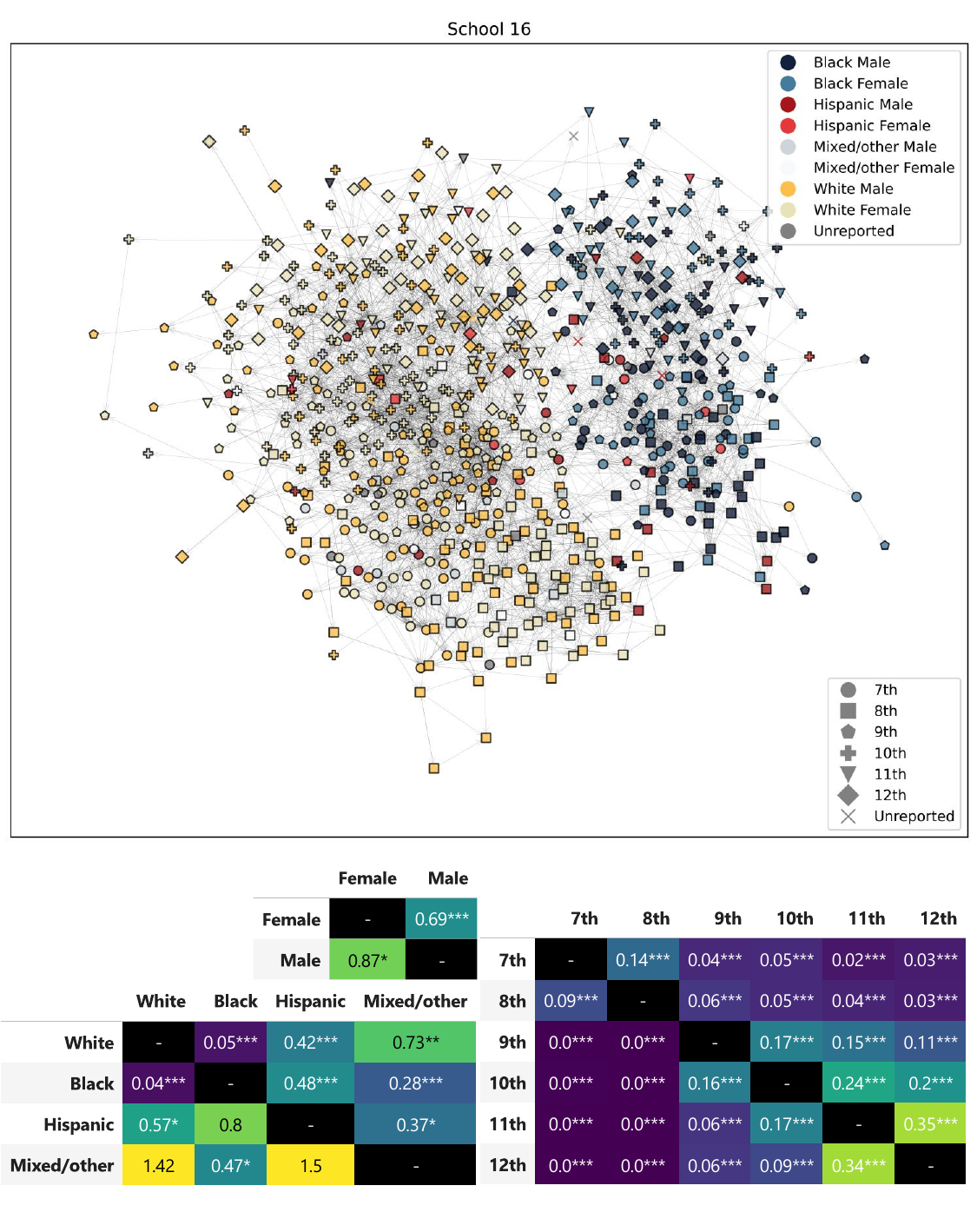}
\caption{Representative example of one of the AddHealth networks and its in-group-normalized latent connection preferences inferred for a 1D+\textbf{AND} model. Asterisks indicate the statistical significance computed using the MRQAP method: $\text{p-value}<0.05^*, 0.01^{**}, 0.001^{***}$.}
\end{figure}

\begin{figure}[h!]
\centering
\includegraphics[width=0.95\textwidth]{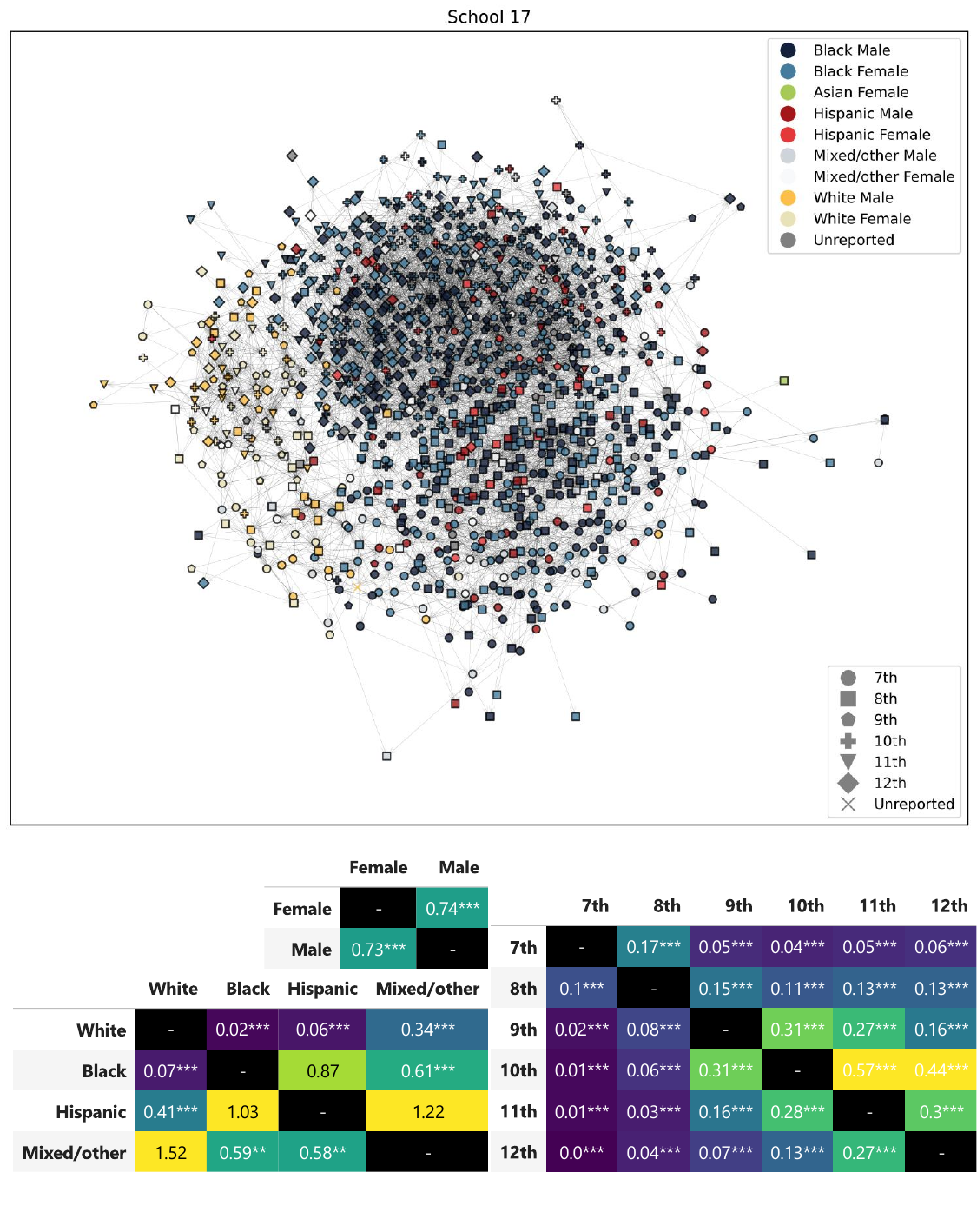}
\caption{Representative example of one of the AddHealth networks and its in-group-normalized latent connection preferences inferred for a 1D+\textbf{AND} model. Asterisks indicate the statistical significance computed using the MRQAP method: $\text{p-value}<0.05^*, 0.01^{**}, 0.001^{***}$.}
\end{figure}

\begin{figure}[h!]
\centering
\includegraphics[width=0.95\textwidth]{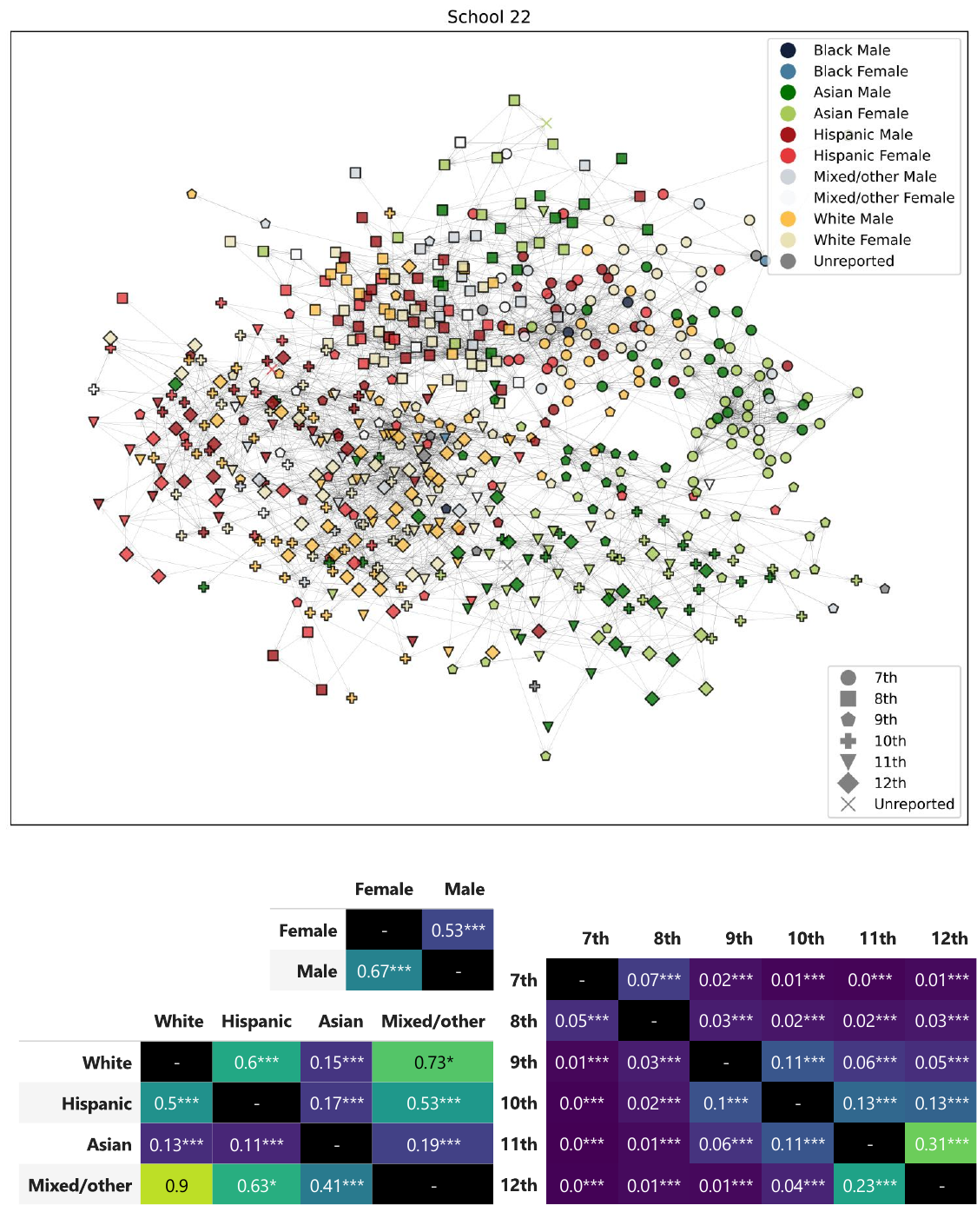}
\caption{Representative example of one of the AddHealth networks and its in-group-normalized latent connection preferences inferred for a 1D+\textbf{AND} model. Asterisks indicate the statistical significance computed using the MRQAP method: $\text{p-value}<0.05^*, 0.01^{**}, 0.001^{***}$.}
\end{figure}

\begin{figure}[h!]
\centering
\includegraphics[width=0.95\textwidth]{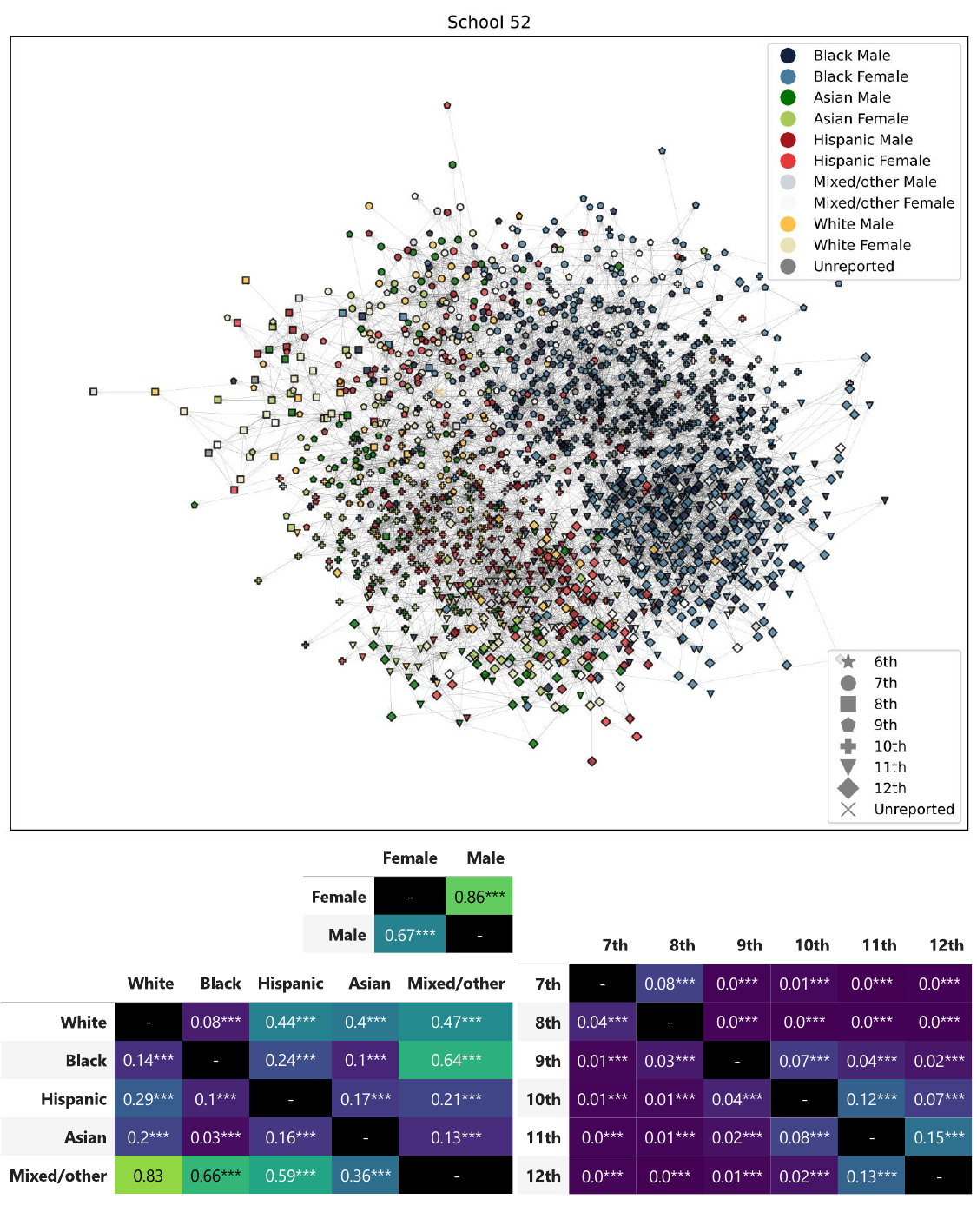}
\caption{Representative example of one of the AddHealth networks and its in-group-normalized latent connection preferences inferred for a 1D+\textbf{AND} model. Asterisks indicate the statistical significance computed using the MRQAP method: $\text{p-value}<0.05^*, 0.01^{**}, 0.001^{***}$.}
\end{figure}

\begin{figure}[h!]
\centering
\includegraphics[width=0.95\textwidth]{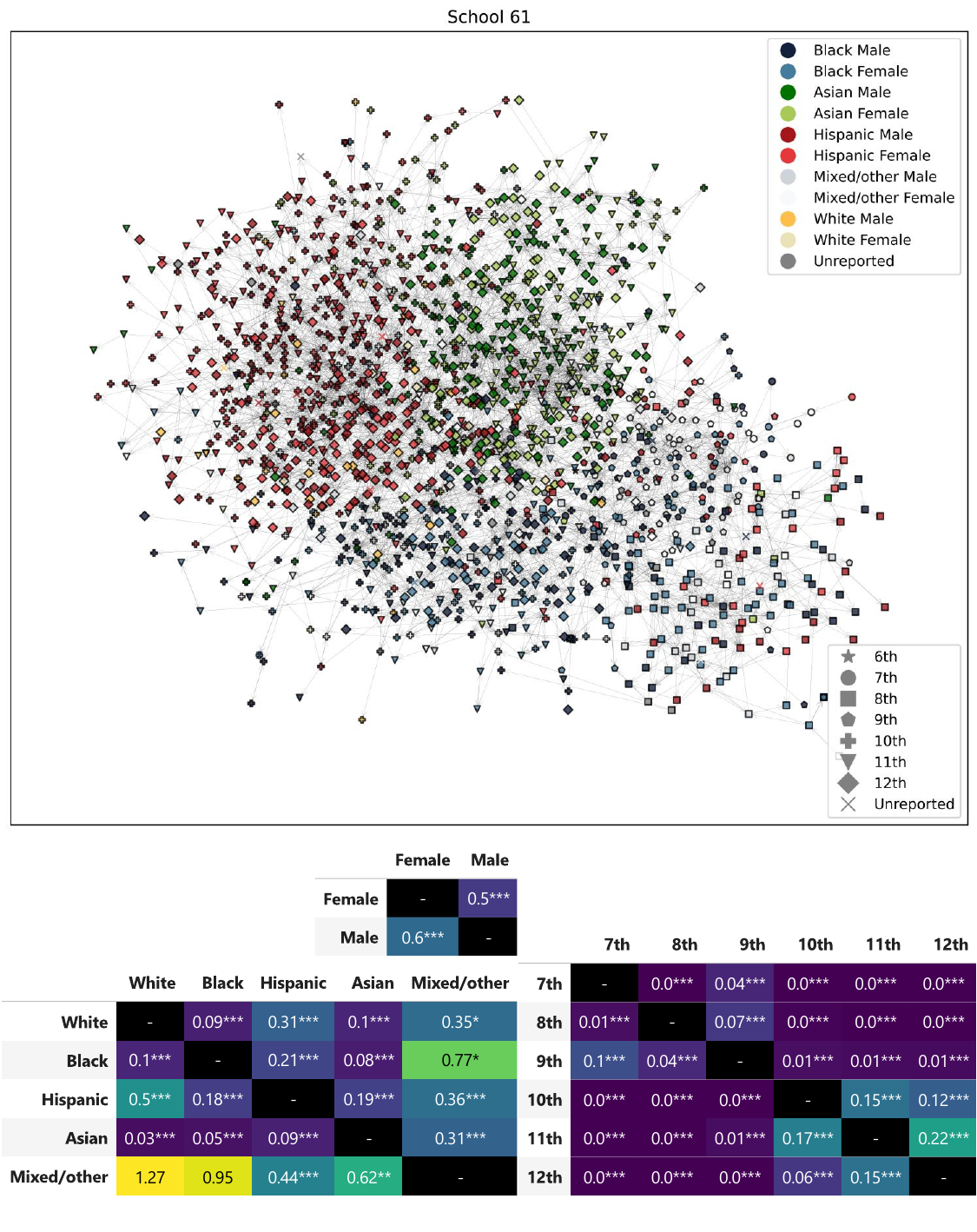}
\caption{Representative example of one of the AddHealth networks and its in-group-normalized latent connection preferences inferred for a 1D+\textbf{AND} model. Asterisks indicate the statistical significance computed using the MRQAP method: $\text{p-value}<0.05^*, 0.01^{**}, 0.001^{***}$.}
\label{fig:example_network_last}
\end{figure}


\end{document}